 \documentclass[12pt,preprint]{aastex}

\newcommand{\myemail}{Georg.Weidenspointner@cesr.fr}
\newcommand{\g}{\ensuremath{\gamma}}

\begin{document}


\title{MGGPOD: a Monte Carlo Suite for Modeling
Instrumental Line and Continuum Backgrounds in Gamma-Ray Astronomy}


\author{G. Weidenspointner\altaffilmark{1,2,3},
M.J. Harris\altaffilmark{1,2}, S. Sturner\altaffilmark{1}, B.J. Teegarden}
\affil{NASA Goddard Space Flight Center, Code 611, Greenbelt, MD 20771, USA}
\email{\myemail}

\and

\author{C. Ferguson}
\affil{Physics Department, University of Southampton, Southampton,
SO17 1BJ, UK}


\altaffiltext{1}{Universities Space Research Association, 7501 Forbes
Blvd., Seabrook, MD 20706-2253, USA} 
\altaffiltext{2}{Centre d'Etude Spatiale des Rayonnements, 9 Avenue
Colonel Roche, 31028 Toulouse Cedex 4, France} 
\altaffiltext{3}{ESA Fellow}


\begin{abstract}

Intense and complex instrumental backgrounds, against which the much
smaller signals from celestial sources have to be discerned, are a
notorious problem for low and intermediate energy \g-ray astronomy
($\sim 50$~keV -- 10~MeV). Therefore a detailed qualitative and
quantitative understanding of instrumental line and continuum
backgrounds is crucial for most stages of \g-ray astronomy missions,
ranging from the design and development of new instrumentation through
performance prediction to data reduction.  We have developed MGGPOD, a
user-friendly suite of Monte Carlo codes built around the widely used
GEANT (Version 3.21) package, to simulate {\sl ab initio} the physical
processes relevant for the production of instrumental
backgrounds. These include the build-up and delayed decay of
radioactive isotopes as well as the prompt de-excitation of excited
nuclei, both of which give rise to a plethora of instrumental \g-ray
background lines in addition to continuum backgrounds. The MGGPOD
package and documentation are publicly available for download from
{\tt http://sigma-2.cesr.fr/spi/MGGPOD/}.  

We demonstrate the capabilities of the MGGPOD suite by modeling high
resolution \g-ray spectra  recorded by the Transient Gamma-Ray
Spectrometer (TGRS) on board {\it Wind} during 1995. The TGRS is a Ge
spectrometer operating in the 40~keV to 8~MeV range. Due to its fine
energy resolution, these spectra reveal the complex instrumental
background in formidable detail, particularly the many prompt and
delayed \g-ray lines.
We evaluate the successes and failures of the MGGPOD package in
reproducing TGRS data, and provide identifications for the numerous
instrumental lines.

\end{abstract}




%%%%%%%%%%%%%%%%%%%%
%   introduction   %
%%%%%%%%%%%%%%%%%%%%

\section{\label{intro} Introduction}

Due to the opacity of the Earth's atmosphere, astronomical
observations in the \g-ray regime must be performed at the top of the
atmosphere or in space, regions that are pervaded by intense radiation
fields. Interactions of the particle radiations with the instrument
and spacecraft materials result in a complex instrumental
background. Discerning the much smaller signals from celestial sources
against this strong background is the single worst problem for \g-ray
astronomy at low and intermediate energies of about 50~keV --
10~MeV. Thus a detailed qualitative and quantitative understanding of
the physical processes giving rise to instrumental line and continuum
backgrounds is crucial for most stages of \g-ray astronomy missions,
ranging from the design and development of new instrumentation through
performance prediction and mission planning to data reduction.

Obtaining quantitative estimates of instrumental backgrounds is
rendered very difficult by the complexity of the physics involved.
The instrumental background of a given mission has a
complicated dependence on its specific radiation environment, which in
turn depends on the mission's orbit and epoch of operation, and on the
details of the instrument design (the detector material, the
field-of-view, active and passive shielding, and the amount and
distribution of active and passive material). For many years semi-empirical
methods and scaling laws have been used to predict the intrumental
background of a given instrument based on data from past missions and
a well-developed understanding of the relevant physics \citep[see
e.g.][]{Dean91, Gehrels92}. However, the subtleties and complexities
involved in scaling from one combination of radiation environment and
instrument design to another degraded the accuracy of these
predictions \citep{Dean03}.

The advent of particle transport codes capable of dependably
representing the fundamental physics involved, together with recent
advances in computer processing speed, make {\sl ab initio} Monte
Carlo simulation a feasible approach for obtaining quantitative
estimates of instrumental backgrounds \citep{Dean03}. In this
approach, model spectra of the mission specific radiation environment
and its time variability are combined with a computer representation
of the geometrical structure and material composition of the \g-ray
instrument and the rest of the spacecraft to track the trajectories
and interactions of the incident particles and their secondaries, and
to record their energy deposits throughout the system. One of the
first Monte Carlo packages developed for this purpose was the
University of Southampton's GGOD suite, which provided the capability
of modeling instrumental line and continuum backgrounds due to the
delayed decay of radioactive isotopes produced in particle
interactions. GGOD has been applied to modeling numerous \g-ray
instruments, among them the GRIS balloon spectrometer, TGRS on board
{\it Wind}, BATSE on board the {\it Compton Gamma-Ray Observatory (CGRO)},
IBIS and SPI on board {\it INTEGRAL}, and BAT on board {\it Swift}
\citep[see][and references therein]{Lei96, Dean03}.  Monte Carlo codes
with similar capabilities have been developed by other groups and were
validated using e.g.\ data from the Oriented Scintillation Specrometer
Experiment (OSSE) on board {\it CGRO} \citep{Dyer94} or from the Ge
spectrometer on board the {\it HEAO-3} mission \citep{Graham97}.

Motivated by the need for accurate modeling of the instrumental line
and continuum background expected for the Ge spectrometer SPI on board
the {\it INTEGRAL} mission, we have improved the GGOD suite and
combined it with the user-friendly NASA/GSFC MGEANT package, which is
e.g.\ used for SPI response simulations \citep{Sturner00,
Sturner03}. Both the MGEANT and GGOD codes are based on the widely
used GEANT Detector Description and Simulation Tool
\citep[Version~3.21,][]{Brun95}. In order to include photons due to
the prompt de-excitation of excited nuclei produced in neutron
captures, inelastic neutron scattering, and spallations in our
simulations, we created the PROMPT package, which is integrated into
the MGEANT and GGOD codes. The resulting suite of Monte Carlo
packages, named MGGPOD, supports {\sl ab initio} Monte Carlo
simulations of both prompt and delayed instrumental backgrounds,
including the plethora of instrumental
\g-ray lines. The MGGPOD suite and documentation are available to the
public for download from {\tt http://sigma-2.cesr.fr/spi/MGGPOD/}. The
various MGGPOD packages, including GEANT Version3.21, have all been
developed in the FORTRAN77 programming language and therefore {\sl
cannot} be used with the more recent Geant4 toolkit, which has been
developed at CERN using the C++ language.

We demonstrate the capabilities of the MGGPOD suite by modeling high
resolution \g-ray spectra recorded by the Transient Gamma-Ray
Spectrometer (TGRS) on board {\it Wind} during 1995 \citep[earlier
modeling using the GGOD suite has been reported by][]{Lei96, Diallo01}. The
TGRS is a Ge spectrometer operating in the 40~keV to 8~MeV range
\citep{Owens95}. Due to its fine energy resolution, these spectra
reveal the complex instrumental background in formidable detail,
particularly the many prompt and delayed \g-ray lines for which we
provide identifications. We evaluate the successes and failures of the
MGGPOD package in reproducing the TGRS line and continuum backgrounds.
A synopsis of the TGRS results presented here, and preliminary
modeling results for the SPI spectrometer, can be found in
\citet{Weidenspointner04a}. Recently, preliminary MGGPOD results have
also been presented for the {\it Reuven Ramaty High-Energy Solar
Spectroscopic Imager} \citep[{\it RHESSI}, described in][]{Smith02} by
\citet{Wunderer04}. A brief overview of the capabilites, functioning,
and structure of the MGGPOD package has been given by
\citet{Weidenspointner04b}.

In this paper, we provide an overview on a variety of aspects
pertaining to the production and characterization of instrumental
backgrounds in \g-ray astronomy in \S~\ref{instrumental_bgd}. The
capabilites, functioning, and structure of the MGGPOD package,
including the modeled physics, are described in \S~\ref{mggpod}. Our
MGGPOD modeling of TGRS data, our TGRS instrumental line
identifications, and detailed comparisons of data and simulation are
given in \S~\ref{mod_tgrs}. A summary of our results and concluding
remarks can be found in \S~\ref{sum-conc}.

%%%%%%%%%%%%%%%%%%%%%%%%%%%%%%%%%%%%%%%
%   Instrumental Background Physics   %
%%%%%%%%%%%%%%%%%%%%%%%%%%%%%%%%%%%%%%%

\section{\label{instrumental_bgd} Instrumental Background}

In this section we provide an overview of a variety of aspects
pertaining to the production and characterization of instrumental
backgrounds in \g-ray astronomy.
The instrumental background of a given \g-ray instrument has a
complicated and complex dependence on its specific radiation
environment and the details of the instrument design. The radiation
environment, i.e.\ the ambient photon and particle radiation fields,
their spectral and angular distributions, as well as their time
histories, are a sensitive function of a mission's orbit relative to
the Earth and the geomagnetic field. The photons and particles that
constitute the local radiation environment interact with detector and
instrument and spacecraft structures through a wide variety of
physical processes, producing secondary photons and particles, as well
as radioactive nuclei, which, upon their (delayed) decay, give rise to
further secondaries. These primary and secondary photons and particles
can result in detector triggers that pass all logical and electronic
criteria required for proper events due to photons from celestial
sources; these triggers constitute the instrumental background. In a
given radiation environment, the strength and spectrum of the
instrumental background are strongly influenced by details of the
instrument design such as the choice of detector material, the
field-of-view, active and passive shielding, and the amount and
distribution of active and passive material.

Among the radiation fields that are potential sources of intrumental
background in \g-ray astronomy missions are cosmic rays (Galactic
cosmic rays, solar energetic particles, and anomalous cosmic rays),
geomagnetically trapped particles, Earth albedo radiations, diffuse
cosmic X and \g\ radiation, and locally produced secondary radiation
\citep[see e.g.][and references therein]{Stassinopoulos89, Klecker96,
Dean03}. An accurate description of the radiation environment clearly
is of great importance for instrumental background
estimates. Table~\ref{rad_env_table} provides a summary (for quick
reference) of some salient features of the radiation background
components described below.
\begin{itemize}
\item {\it Galactic cosmic rays}: Galactic cosmic rays primarily consist of
protons (about 85\% by number), followed by $\alpha$ particles ($\sim
12$\%), and electrons ($\sim 2$\%); the remainder are heavier
nuclei. For kinetic energies above about 10~GeV/nucleon solar
modulation is no longer effective and the differential energy spectrum
of cosmic rays can be approximated by an $E^{-2.7}$ power law.
This particle radiation is a dominant source of
instrumental background for missions that spend most of their time
outside the Earth's magnetosphere or at least above the radiation
belts, but may be less significant for balloon borne experiments or
missions in low-Earth orbits (LEOs) because the geomagnetic field
acts as a momentum/energy filter requiring that a charged particle
must exceed a minimum rigidity\footnote{The rigidity $R$ of a particle
of charge $q$ and momentum $p$ is given by: $R =
\frac{p c}{q}$. A cut-off rigidity of 10~GV, which is a ``typical''
value for LEOs with low inclination, implies that a cosmic-ray proton
must exceed a kinetic energy of about 9.1~GeV to reach the spacecraft}
(the so-called cut-off rigidity) to reach a given location within the
magnetosphere. For LEOs the incident cosmic radiation therefore varies
with time because the local cut-off rigidity changes as the spacecraft
follows its trajectory. Outside the magnetosphere the only source of
temporal variation in the incident cosmic radiation is solar
modulation, which affects particles with kinetic energies less than a
few GeV/nucleon. At balloon altitudes and in LEO the geomagnetic
cutoff and Earth shadowing result in a highly anisotropic cosmic-ray
intensity; outside the magnetosphere cosmic rays can be assumed
isotropic. 
\item {\it Solar energetic particles and anomalous cosmic rays}: Solar
energetic particles (SEPs) are produced in violent energy releases on
the Sun such as flares or coronal mass ejections (CMEs). They are
reinforced on their way to Earth by interplanetary particles
accelerated by the shock waves from CMEs. During these times their
flux can be enormously high compared to other particle
radiations. SEPs have energies up to several hundred MeV/nucleon and
hence a lower rigidity than Galactic cosmic rays. Missons in LEOs and
balloon experiments are therefore shielded from most of them, while
SEPs can be a dominant temporary source (up to a few days) of
instrumental background for missions outside the magnetosphere. Like
SEPs, anomalous cosmic rays are mainly a concern for missions outside
the magnetosphere; because they consist mostly of heavy nuclei they
can be effective producers of secondaries in nuclear interactions. The
intensity of anomalous cosmic rays is much more strongly effected by
the solar cycle than that of Galactic cosmic rays.
\item {\it Geomagnetically trapped radiation}: Out to a few Earth
radii, the geomagnetic field can be well approximated by a
dipole. Charged particles can be magnetically trapped in such a dipole
field and stored in so-called radiation belts in the geomagnetic
equatorial plane. The most abundant trapped particle species are
protons and electrons. Protons, which are of main concern for
instrumental backgrounds, are trapped in a single belt, with the
maximum proton intensity occuring at an altitude of $\lesssim 1$~Earth
radius. The energies of the protons extend up to several hundred MeV;
their angular distribution can be highly anisotropic
\citep{Whatts89}. The geomagnetic dipole is offset from the Earth's
center, therefore the proton belt appears to extend to lower altitudes
over the coast of Brazil -- this area is usually referred to as the
South Atlantic Anomaly (SAA). Because of its intensity this trapped
proton radiation is avoided as much as possible by \g-ray
missions. However, the SAA usually can not be avoided altogether for
missions in LEOs. If encountered, passages through the SAA are a
dominant source for the production of radionuclides in the instrument
and spacecraft materials
\citep{Kurfess89, Weidenspointner01}. The SAA dosage received by a
mission in LEO strongly depends on the orbit's altitude and
inclination.
\item {\it Earth albedo radiation}: When entering the Earth's atmosphere
Galactic cosmic-ray particles interact violently with the air nuclei,
initiating nuclear interaction cascades that ultimately result in the
production of a multitude of secondaries of relatively low energy. The
most important of these secondaries for \g-ray experiments at balloon
altitudes and in LEO are photons and neutrons. At balloon altitudes,
the angular distribution of these secondaries is highly anisotropic,
and their intensity depends on the local geomagnetic cut-off rigidity
and on the depth in the atmosphere \citep[see e.g.][and references
therein]{Gehrels85}. Depending on energy, both albedo photons and
neutrons can constitute strong and anisotropic radiation fields for
missions in LEO, and are dominant background sources for balloon
experiments. The relative importance of the Earth's \g-ray albedo as a
diffuse photon source increases with energy, above about 35~MeV it is
10--100 times more intense than the diffuse cosmic \g\ radiation
\citep{Thompson_Simpson81}. The energy distribution of albedo neutrons
extends from thermal energies to at least several hundred MeV.  The
intensity of albedo \g-rays \citep{Thompson_Simpson81, Harris03} and
neutrons \citep{Weidenspointner96, Morris98} varies with the local
geomagnetic rigidity and with the solar cycle.
\item {\it Diffuse cosmic X and \g\ radiation}: The diffuse cosmic X and \g\
radiation is of great astrophysical interest in its own right
\citep[e.g.][]{Weidenspointner_Varendorff01}; however, it constitutes
a background against which all other observations must be made. For
the purpose of background simulations, the diffuse cosmic X and \g\
radiation can be considered isotropic \citep{Kinzer97, Sreekumar98,
Weidenspointner00}, and its energy spectrum constant in time and
unaffected by the geomagnetic field. For missions in LEO the shielding
effect of the Earth needs to be taken into account. In addition, at
balloon altitudes there is direction-dependent attenuation by the
atmosphere, although the atmospheric \g\ radiation usually dominates.
\item {\it Internally produced secondary particles}: The photons and
particles of the external radiation fields described above interact
with the instrument and spacecraft structures through a wide variety
of physical processes (see below), producing a multitude of secondary
particles such as photons, electrons and positrons, neutrons, protons
and heavier nuclei. Because this locally produced secondary radiation
is produced by the external radiation fields it is therefore not a
component of the radiation environment in the same sense as the
external components. We list here the secondary radiation as a
separate component of the radiation environment because it is treated
as such in semi-empirical calculations of instrumental backgrounds
\citep[e.g.][]{Gehrels92}. In {\sl ab initio} Monte Carlo
calculations, the topic of this paper, these secondary particles are
derived from the modeling of the interactions of the
external radiation fields, which are the only input.
\end{itemize}

The photons and particles of the radiation environment interact
through a wide variety of processes with the instrument and spacecraft
structures and contribute to the instrumental background in different
ways. Backgrounds due to external photon fields are most important for
instruments with poor spatial resolution and a large
field-of-view. The photons trigger the detector either by entering
through the field-of-view (aperture flux) or by entering through the
veto shield without it being triggered (shield leakage).  Energetic
protons (and heavier nuclei) are a major source of instrumental
background. They may pass through the instrument and spacecraft
suffering only slight energy losses due to ionization in the traversed
materials. However, they may also undergo a catastrophic nuclear
interaction with a single nucleus, creating a potentially large number
of secondary particles such as pions, nucleons, and light
nuclei; usually a relatively heavy product nucleus is left behind.
The secondary particles may be energetic enough to initiate further
nuclear interactions with other nuclei, their decay (e.g.\ pions) can
result in the generation of secondary photons, electrons, and
positrons, and the de-excitation of residual nuclei gives rise to more
secondary particles (mainly photons).

All of these processes and interactions occur on timescales that are
short compared to typical timescales for event processing by the
instrument electronics (which typically are a few to about 100~$\mu$s)
and contribute to the so-called {\it prompt} instrumental
background. Depending on energy, veto shields can reduce prompt
background components by a few orders of magnitude. Some of the
product nuclei resulting from interactions of protons or other hadrons
may not be stable and emit further secondary particles when undergoing
radioactive decay. Depending on the lifetime of the unstable product
nucleus, the time delay between its production and its decay may be
much longer than typical instrumental event processing timescales,
contributing to the so-called {\it delayed} instrumental
background. For satellite missions the long-term build-up of delayed
background due to long-lived isotopes (e.g.\ $^{22}$Na, $T_{1/2} =
2.7$~y) is very clearly visible \citep{Kurfess89,
Weidenspointner01}. The best strategy to minimize delayed instrumental
background is to minimize the amount of passive material in the
instrument, particularly in the vicinity of the detectors. More
positively, the time-scale of the long-term build-up is characteristic
of the nuclear half-life, and a valuable key to identification of the
nucleus involved.

Neutrons are another major source of both prompt and delayed
instrumental background components. Unlike protons, neutrons can
traverse veto shielding without triggering it and produce prompt
background {\sl within} the detector through elastic and inelastic
scattering (by passing energy to the recoil nucleus and de-excitation)
as well as neutron capture (de-excitation of the product
nucleus). Outside the veto shield the de-excitation of product nuclei
from neutron induced nuclear interactions, including neutron capture,
also contribute to the prompt background. Some of these product nuclei
are unstable, and their radioactive decay adds to the delayed
background.

An important aspect of the instrumental background is the presence of
many lines. The position, strength, and shape of instrumental lines
are relevent because they can interfere with spectroscopy of
astrophysical \g-ray lines. Instrumental lines can result from prompt
processes such as de-excitation of excited nuclei as well as from
delayed radioactive decays. Instrumental lines can also be used as
powerful diagnostics of the overall instrumental background, including
continuum components. For example, \g-ray lines produced by isomeric
transitions in Ge detectors can be used to estimate the neutron flux
inside the veto shield, which then in turn allows an estimate of
continuum backgrounds due to elastic and inelastic neutron scattering
and neutron induced $\beta$-decays
\citep[e.g.][]{Naya96}.

%%%%%%%%%%%%%%
%   MGGPOD   %
%%%%%%%%%%%%%%

\section{\label{mggpod} The MGGPOD Monte Carlo Simulation Suite}

The MGGPOD suite is a user-friendly Monte Carlo simulation package
that is applicable to all stages of space-based \g-ray astronomy
missions.  In particular, the MGGPOD suite allows {\sl ab initio}
simulations of instrumental backgrounds -- including the many \g-ray
lines -- arising from interactions of the various radiation fields
within the instrument and spacecraft materials. It is possible to
simulate both prompt instrumental backgrounds, such as energy losses
of cosmic-ray particles and their secondaries, as well as delayed
instrumental backgrounds, which are due to the decay of radioactive
isotopes produced in nuclear interactions. MGGPOD can also be used to
study the response of \g-ray instruments to astrophysical and
calibration sources. The MGGPOD suite is therefore an ideal Monte
Carlo tool for supporting most stages of \g-ray missions, ranging from
%design, development, and performance prediction through calibration
%and response generation to data reduction; 
instrument design to data reduction. Software and documentation are
available to the public for download at the Centre d'Etude Spatiale
des Rayonnements\footnote{http://sigma-2.cesr.fr/spi/MGGPOD/}. In this
publication we focus on the physics simulated by the MGGPOD
suite. Detailed practical advice for users on how to install and use
this Monte Carlo package, including examples, can be found in the
documentation available on the MGGPOD web site \citep[see
also][]{Weidenspointner04b}.

MGGPOD is a suite of five closely integrated Monte Carlo packages,
namely {\bf MG}EANT, {\bf G}CALOR, {\bf P}ROMPT, {\bf O}RIHET, and
{\bf D}ECAY, each of which will be described in more detail below. The
MGGPOD package resulted from a combination of the NASA/GSFC MGEANT
\citep{Sturner00, Sturner03} and the University of Southampton's GGOD
\citep{Lei96, Dean03} Monte Carlo codes. Both were improved, and 
supplemented by the newly developed PROMPT package. The overall
structure of the MGGPOD package is illustrated in
Fig.~\ref{mggpod_flow_chart}.  Depending on the simulated radiation
field or \g-ray source distribution one or three steps, requiring two
or three input files, are needed to obtain the resulting energy
deposits in the detector system under study, as summarized in
Table~\ref{mggpod_sim_steps}. In general, it is advisable to simulate each
component of the radiation environment separately. MGGPOD
distinguishes two classes of radiation fields. Class~I comprises
radiation fields for which only prompt energy deposits are of
interest, such as celestial or laboratory \g-ray sources or cosmic-ray
electrons. Class~II comprises radiation fields for which additional
delayed energy deposits resulting from the activation of radioactive
isotopes need to be considered. Examples of Class~II fields are
cosmic-ray protons, or geomagnetically trapped protons.

For both of these classes, the simulation of the prompt energy
deposits requires two input files: a mass model, and a model of the
simulated radiation field. The mass model is a detailed computer
description of the experimental set-up under study. It specifies the
geometrical structure of instrument and spacecraft, the atomic and/or
isotopic composition of materials, and sets parameters that influence
the transport of particles in different materials. Each component of
the radiation environment (and analogously for \g-ray sources) to
which the instrument is exposed is characterized by three quantities:
the type of the incident particles, and their spectral and angular
distributions. The prompt energy deposits are written to an output
event file. 
%For a Class~I radiation field MGEANT is sufficient for the
%simulation, for a Class~II radiation field an executable linking the
%MGEANT, GCALOR, and PROMPT packages is required. 
In case of a Class~II
radiation field there is an additional output file which lists, for all
nuclei produced in hadronic interactions, the product nucleus'
identity along with the geometrical mass model element in which it was
produced. From this file isotope production rates can be computed.

To simulate delayed energy deposits (Class~II radiation field) two
additonal steps need to be taken. These require as input the time
history of the radiation field which is
responsible for the activation, and the previously calculated isotope
production rates.
Based on this information first the activity of each isotope produced
in each structural element of the mass model is determined by
ORIHET. Then, employing the MGEANT, GCALOR, and DECAY packages, these
activities are used to simulate the delayed energy deposits due to
radioactive decays in the instrument.

Combining prompt and delayed energy
deposits from each component of the radiation environment and
\g-ray sources, it is possible to obtain the total energy deposited
in the system as a function of position and time.
In the following, each of the five packages that constitute the MGGPOD
simulation suite is described.

\notetoeditor{Fig. 1 should be placed at the top of a page, with a
width equal to the width of the text on the page.}

\subsection{\label{mgeant} MGEANT}

MGEANT is a multi-purpose simulation package developed by the Low
Energy Gamma Ray Group (LEGR) at NASA/GSFC \citep{Sturner00,
Sturner03}. It is based on the GEANT Detector Description and
Simulation Tool (Version 3.21) created and supported by the
Application Software Group, Computing Networks Division, at CERN
Geneva, Switzerland \citep{Brun95}. GEANT3 is designed to simulate the
passage of elementary particles through an experimental set-up, which
may be of considerable complexity.  Although originally designed for
high-energy physics experiments, GEANT3 has found applications in many
other areas, including space science and specifically \g-ray
astronomy. Within the MGGPOD suite, MGEANT (i.e.\ GEANT3) stores and
transports all particles, and treats electromagnetic interactions from
about 10~keV to a few TeV. In addition, MGEANT provides the option to
use the GEANT Low-Energy Compton Scattering package GLECS, which
provides more detailed physical models of the coherent (Rayleigh) and
incoherent (Compton) photon scattering processes than those included
in the standard GEANT3 distribution by taking into account the kinetic
energy of the bound electrons \citep{Kippen02,
Kippen04}\footnote{Version~1.1 of MGGPOD, which is currently under
development, will in addition provide the option to use the GLEPS
package to take into account the polarization of the initial photon in
Compton and Rayleigh scatterings \citep{Kippen02}.}.

MGEANT was created to increase the versatility and user-friendliness
of the GEANT3 simulation tool\footnote{MGEANT does not work with the
newly developed, C++ based, Geant4 toolkit.}. A modular, ``object
oriented'' approach was pursued, giving MGEANT two main advantages
over standard GEANT3. First, the instrument specific geometries and
materials are provided via input files, rather than being
hard-coded. Second, several event-generation beam models and spectral
models are available \citep[e.g.][]{Sturner00, Sturner03}. Beam models
include a plane wave, an isotropic radiation field, astrophysical and
calibration point sources, or a user defined sky map; spectral models
include power law and exponential spectra, line emission, or user
defined spectra. Furthermore, MGEANT can interactively display (using
CERN's PAW++ package) the geometric set-up as well as the particle
trajectories -- a very convenient capability when creating a mass
model or defining and verifying beam parameters.  In MGGPOD the only
supported output file format is FITS, for which two different event
list formats are available: 1) standard format where the total energy
deposit in each detector (and in the anti-coincidence system) for each
event is listed, and 2) extended format where the energy deposits and
location of each interaction, the time of interaction, and the type of
the interacting particle are also listed. The extended format was
introduced to facilitate detailed instrument response studies, such as
the pulse-shape discrimination (PSD) system of the {\it INTEGRAL}
spectrometer SPI, or event reconstruction algorithms for advanced
Compton telescope concepts.

MGEANT is therefore very well suited for rapid prototyping of detector
systems, it can readily generate most of the radiation fields relevant
to \g-ray astronomy, and it is set up to support detailed instrument
response studies. The MGEANT simulation package and a user manual are
available at the NASA/GSFC web site of the LEGR group\footnote{\sl
http://lheawww.gsfc.nasa.gov/docs/gamcosray/legr/mgeant/mgeant.html}.

\subsection{\label{gcalor} GCALOR}

GCALOR \citep{Zeitnitz_Gabriel94, Zeitnitz_Gabriel99} is an interface
between the CALOR89 package by \citep{Gabriel95} and the GEANT3 simulation
tool. The CALOR89 package, designed to simulate calorimeter systems
for high-energy physics detectors, simulates hadronic interactions
down to 1~MeV for nucleons and charged pions and down to thermal
energies ($10^{-5}$~eV) for neutrons.
GCALOR incorporates the capability to perform hadronic interaction
calculations in the GEANT3 framework by extracting the HETC (High Energy
Transport Code) collision and evaporation model and the FLUKA model
(which is already available in recent versions of GEANT3) from
CALOR89. The HETC Monte Carlo code consists of two parts: the
Nucleon Meson Transport Code (NMTC) and the Scaling Model (which
provides a smooth transition between the NMTC and FLUKA regimes).
For low energy neutrons the MICAP (Monte Carlo Ionization
Chamber Analysis Package) neutron code by \citet{Johnson_Gabriel88} has
been included into GCALOR, rather than the MORSE neutron transport
code utilized in CALOR89\footnote{The performance of MORSE is very
similar to that of MICAP \citep{Johnson_Gabriel87}. However, MICAP is
easier to interface with GEANT3 than MORSE \citep[][Zeitnitz 2004,
priv.\ comm.]{Zeitnitz_Gabriel94}. The MORSE program uses an
energy binning for neutron cross-sections and secondary energy
distributions. Neutrons are generated only with these discrete
energies. MORSE does account for particle multiplicities by assigning
a weight to the generated secondaries. In contrast, MICAP uses
pre-processed ENDF/B cross-section data and the number of energy
points is not fixed, but depends on the isotope and the availability
of cross-section data for this isotope. MICAP statistical weights for
secondary particles are substantially smaller than those of MORSE,
hence it is easier to generate discrete numbers of actually produced
secondary particles which are then handed to GEANT3, which cannot
treat statistical weights, for further processing.}.
The particle types and energy ranges covered by the four modules of
GCALOR are:
\begin{itemize}
\item {\bf NMTC}: nucleons 1~MeV to 3.5~GeV, charged pions 1~MeV
to 2.5~GeV
\item {\bf Scaling Model}: nucleons and charged pions 3~GeV to 10~GeV
\item {\bf FLUKA}: nucleons and charged pions above 10~GeV, and for all
energies for particle types not implemented in CALOR
\item {\bf MICAP}: neutrons $10^{-5}$~eV to 20~MeV
\end{itemize}
Hadronic interactions between particles that are both heavier than
individual nucleons, such as deuterium or helium nuclei, cannot be
simulated with GCALOR. However, continuous energy losses, for example
due to ionization of the matter traversed, are taken into account for
all charged particles. The GCALOR package is publicly available at the
Universit{\"a}t
Mainz\footnote{http://wswww.physik.uni-mainz.de/zeitnitz/gcalor/gcalor.html}.

When combining GCALOR with GEANT3 (or MGEANT), all particles are stored
and transported by GEANT3. Electromagnetic interactions are simulated
by GEANT3, hadronic interactions are simulated by GCALOR. The GEANT3
plus GCALOR package therefore extends the capabilities of the standard
GEANT3 tool to include hadronic interactions of charged particles down
to 1~MeV and of neutrons down to thermal energies. Equally important,
this package provides access to the energy deposits from all
interactions as well as to isotope production anywhere in the
geometrical set-up.

Originally, GCALOR/MICAP utilized only ENDF/B (Evaluated Nuclear
Data Files version B) neutron data.
Unfortunately, these data, available e.g.\ at
the National Nuclear Data Center (NNDC) at Brookhaven National
Laboratory\footnote{\sl http://www.nndc.bnl.gov/}, do not cover all
elements or isotopes. In particular, individual neutron cross-sections
for the five natural Ge isotopes are missing. These cross-sections,
which are clearly of great importance for simulations of instrumental
backgrounds in Ge spectrometers, were generated based on JENDL
(Japanese Evaluated Nuclear Data Library\footnote{\sl
http://wwwndc.tokai.jaeri.go.jp/jendl/jendl.html}) data by Zeitnitz
(2001, priv.\ comm.).

Some of the low-mass residual nuclei produced in spallations as
modelled by GCALOR are very neutron rich, particularly if the incident
projectile is energetic (e.g.\ a primary cosmic-ray proton). Most of
these neutron-rich nuclei result from a failure of the intranuclear
cascade/evaporation model used in GCALOR (Gabriel 2004, priv.\
comm.). Some neutron rich nuclei have been identified in
nuclear experiments and found to have short lifetimes of less than
1~s. Neutron rich nuclei are unstable against the emission of one
or several neutrons, a process that affects the number and energy
distribution of secondary neutrons. To include this source of
secondary neutrons into our simulations, MGGPOD checks whether light
product nuclei ($Z \leq 8$) are unstable against neutron emission,
exploiting existing data whenever available
\citep{Firestone96}, and mandates neutron emission if appropriate,
converting the unstable original product nucleus into the highest-mass
isotope that has no neutron exit channel.

\subsection{\label{prompt} PROMPT}

The PROMPT package was created for simulating the prompt de-excitation
of excited nuclei produced by neutron capture, inelastic neutron
scattering, and spallation\footnote{Although spallation is physically
defined in terms of the evaporation model (see
\S~\ref{prompt_spallation}), for the purpose of MGGPOD it is defined
to comprise all hadronic interactions other than neutron capture which
result in a product nucleus that is different from the target
nucleus. We will, however, assume the validity of the evaporation
model for all such interactions in our treatment}.  The PROMPT package
consists of a data base containing information on the de-excitation of
a large number of nuclei, and of code to access these data and to
generate random samples of de-excitation particles (photons and
electrons). When simulating prompt hadron-induced backgrounds with
MGGPOD, the MGEANT, GCALOR, and PROMPT packages are linked into a
single executable (see Fig.~\ref{mggpod_flow_chart}). Each time a
secondary nucleus is produced in a nuclear interaction as simulated by
GCALOR, PROMPT is called to model the de-excitation cascade. In
GCALOR, prompt photons are generated for a few hadronic interactions,
however, in general these photons are ``statistical'' and continuous
-- they do not reflect the actual discrete nuclear levels and the
well-known selection rules for transitions between them. This
approximate treatment proved sufficient for modeling calorimeters for
high-energy physics detectors \citep{Zeitnitz_Gabriel94}, but is
clearly insufficient for detailed modeling of instrumental \g-ray
backgrounds. Therefore any prompt photons generated by GCALOR are
replaced with those returned by the PROMPT package. With the exception
of a few selected neutron rich nuclei (see previous \S), PROMPT treats
all secondary nuclei as stable, irrespective of their lifetime; all
radioactive decays are treated in a subsequent simulation step using
the DECAY package (see Fig.~\ref{mggpod_flow_chart} and
Table~\ref{mggpod_sim_steps}). A de-excitation cascade may proceed
through one or more isomeric levels; if so, their effect on the time
sequence of de-excitation particle emission is taken into account
approximately, as described in \S\ref{decay}.

We developed the PROMPT package in the spirit of the compound nucleus
model, in which it is assumed that the reaction process can be
separated into independent incoming and outgoing channels with a
well-defined intermediate compound state.  We rely on the GCALOR
package for a complete specification of the incoming channel in the
three classes of prompt reaction, i.e.\ neutron capture, inelastic
neutron scattering, and spallation.  Beyond a determination of the
daughter nucleus, the outgoing channel is, however, at best
incompletely specified by GCALOR; in particular, de-excitation photons
are either missing or incorrect, as described above.  In other words,
GCALOR generates a known daughter nucleus in an unknown excited state
with unknown de-excitation properties. In addition, the distribution
of excited states left behind by the three types of prompt reaction is
not generally known -- unlike the case of radioactive decay, where the
excited state in the daughter has a well-known probability from
extensive measurements of branching ratios (see \S~\ref{decay}).  To
create the PROMPT package we therefore had to derive (simple) recipes
for selecting an excited state in the daughter nucleus and then
modeling its de-excitation into known lower energy levels.  Our
approach is to use a probabilistic recipe to specify the intermediate
excited state of the compound daughter nucleus, and a further {\em
ansatz\/} to specify the de-excitation of the excited state into known
lower levels.  When generating de-excitation particles (usually
photons), we do not take into account Doppler broadening due to the
motion of the nucleus during de-excitation, which is relevant for
short-lived levels with $T_{1/2} \la 0.5$~ps \citep{Evans02}.

\subsubsection{Neutron Capture and Inelastic Neutron Scattering
\label{prompt_ncap_insct}}

In the case of neutron capture and inelastic neutron scattering, the
excitation energies involved are fairly low, and the processes
involved are fairly well understood.  We selected the excited state of
the compound nucleus at random from a level density formula, by which
the probability of obtaining an energy level depends on its spin $J$,
parity $\pi$ and excitation energy $U$.  The value of $U$ was obtained
from GCALOR from the kinematics of the incoming channel reaction.  The
distribution of spins and parities follows the angular part of the
Bethe level density formula as parametrized by
\citet{Mughabghab_Dunford98}; this angular part contains a weak
dependence on $U$ which we neglected (the overall density contains an
exponential in $\sqrt{U}$, but we are only interested in the relative
probabilities of different $J$ and $\pi$ values). Having selected an
excited state in this way, we assumed that it coupled to known states
of lower excitation energy (from the ENSDF database\footnote{Evaluated
Nuclear Structure Data Files (ENSDF) are available at the NNDC (see
footnote~11).}) by an electric dipole (E1) transition -- the spins and
parities of the lower states were required to obey the well-known E1
selection rules for coupling to $J^{\pi}$, and if none were permitted
we assumed magnetic dipole (M1) transitions.  The probability of a
transition from an excited state to a permitted level $x$ was assumed
to be proportional to $(U - U_{x})^3$, which holds approximately for
dipole transitions \citep{Blatt_Weisskopf52}. The specific
prescriptions for generating prompt photons for these two
neutron-induced reactions are outlined in the following, and
illustrated in schematic form in Figs.~\ref{ncap_scheme} and
\ref{insct_scheme}. 

For neutron captures on nuclei with $A>3$ (M)GEANT and GCALOR specify
the kinetic energy of the incident neutron, the target and the product
nucleus, and the recoil energy $E_r$ of the product nucleus. We
estimated the excitation energy $U$ of the product nucleus as the sum
of its neutron separation energy $S_n$ \citep[taken from][typically
several MeV]{Firestone96}, which is the excitation energy after
capturing a thermal neutron, and the kinetic energy of the incident
neutron $E_k$; the recoil energy of the product nucleus typically is
only a few eV and therefore negligible.  We obtain the probability of
a compound nucleus level having given values of $J^{\pi}$ from the
\citet{Mughabghab_Dunford98} level density formula as described above;
however, we make the approximation that it is always evaluated at
excitation energy $U = S_{n}$, which is justified by the weak
variation of the $J^{\pi}$ distribution on $U$.  The de-excitation
cascade which follows was treated by two different {\em ans\"{a}tze},
depending on the value of $J^{\pi}$; those values which are compatible
with thermal neutron capture were simply assigned the branching ratios
from the extensive thermal neutron database maintained by NNDC, while
all other $J^{\pi}$ values were assumed to de-excite by the E1
mechanism described above.  Thermal neutrons couple to the target
nucleus by low partial waves in the incoming channel (we assumed
s-wave), so that the corresponding $J^{\pi}$ in the compound nucleus
are easily specified, given the target nucleus' ground state. For
example, in the common case of an even-even target ($0^{+}$), an
s-wave corresponds to $\frac{1}{2}^{+}$ in the compound nucleus.  Our
motivation was to exploit the very large amount of 
thermal neutron experimental data available in this special case.  The
product nuclei for which the prompt de-excitation cascade after
neutron capture is modeled in these ways are listed in
Table~\ref{n_isot_table}. 

For inelastic neutron scattering on nuclei with $A>4$ (M)GEANT and
GCALOR specify, among other quantities, the kinetic energy of the
incident neutron $E_k$, the recoil nucleus and its kinetic energy
$E_r$, as well as that of the scattered neutron $E_k^\prime$. The
excitation energy of the recoil nucleus $U$ is assumed to be $U = E_k
- E_k^\prime - E_r$. 
In general, the GCALOR cross-sections are accurate enough to reflect
the lowest levels. However, above about 1--2~MeV the derived excitation
energies approach a ``continuum''. Modeling of the de-excitation
cascade depends on the excitation energy $U$. If it is less than or
equal the energy of the highest known level with transition data in
ENSDF, $E_h$, then the initial level of the de-excitation cascade is
chosen to be the level with transition data whose energy is closest to
$U$. If $U$ is greater than $E_h$ then we assume a single initial
transition from $U$ down to any of those known levels for which
spin-parity and transition probability data exist.
The probability for this initial transition is calculated from the
standard spin-parity distribution and the dipole selection rules as
indicated above. The spin-parity distribution is evaluated at a
reference energy $E_{\mathit{ref}}$ chosen for each nucleus to be
approximately where the inelastic neutron scattering cross-section
peaks according to JENDL, which characteristically is around
10~MeV. As in the case of neutron capture (where we used $U = S_{n}$),
we assume that the calculated probability distribution for the initial
transition from $E_{\mathit{ref}}$ applies to all excitation energies
$U$ greater than $E_h$. This is justified by the weak dependence of
this distribution on $U$. In the de-excitation cascade from this
level at excitation energy $U$ the energy of the initial transition is
increased by $U-E_{\mathit{ref}}$. The recoil nuclei for which the
prompt de-excitation cascade after inelastic neutron scattering is
modeled are listed in Table~\ref{n_isot_table}. 

The existence of asymmetrically broadened triangular or
``sawtooth'' shaped features extending in energy above the lines from
the lowest energy neutron inelastic scatters has been recognized by
several authors in both laboratory
\citep{Bunting_Kraushaar74}
and flight \citep{Wheaton89, Evans02, Weidenspointner03} data, and
are seen in our data also. They arise from the summed energy
depositions by the recoiling Ge nucleus and by the de-excitation
photon(s). These features are automatically predicted by (M)GEANT and
GCALOR and furnish a comparison between the simulation and the
behavior of the spectrometer.

\subsubsection{\label{prompt_spallation} Spallation}

For spallation reactions (M)GEANT and GCALOR specify the type and
kinetic energy and momentum of the incident particle, the target
nucleus, and the types, energies and momenta of all reaction products.
However, because GCALOR conserves energy and momentum for any
interaction only on average, but not rigorously for an individual
case, it is not possible to obtain a meaningful estimate of the
excitation energy $U$ of any reaction product from kinematics.  Nor
are experimental cross-sections available for production of such
individual levels, which are extremely numerous.  However, some
general rules about the distribution of $U$ values follow from the
evaporation model, which is the generalization of the compound nucleus
model to particle emission \citep{Weisskopf_Ewing40}.  Proton-induced
spallation reactions at high energy leave the compound nucleus in a
very broad distribution of excitation energies, which can be thought
of as a thermal distribution $\sim \exp(-U/kT)$ for some temperature
parameter T \citep[unlike heavy-ion collisions, which tend to select
collective states such as the high-spin yrast levels:][]{Galin01}.
If $T$ exceeds the nucleon separation energy one or more nucleons
are likely to be emitted (neutrons above $S_{n}$ being favored, since
there is no Coulomb barrier).  The generation of a daughter nucleus by
GCALOR implies that neutron emission has ceased, which in turn implies
that $T < S_{n}$.  Characteristic values of $S_{n}$ are several MeV;
the fact that the thermal distribution has a substantial ``tail'' at
energies above $T$ implies that neutron emission only ceases when $T$
falls far enough below $S_{n}$ for the ``tail'' to become negligible.
Experiments attempting to achieve high $T$ in residual nuclei left
behind by proton reactions on nuclei between Ag and Au have attained
up to $T \simeq 5$~MeV, with values 3--4~MeV being characteristic
\citep{Ledoux98}.  On this basis we assumed that the known excited states
$U_{x}$ in spallation product nuclei were distributed according to the
formula $(2J_{x} + 1) \exp(-U_{x}/3~{\rm MeV})$. A sketch of the
procedure followed once particle emission has stopped is given in
Fig.~\ref{spal_scheme}. 

A key assumption behind this type of argument is that nuclear energy
levels can be treated statistically, which is recognised to become 
less and less valid as the nuclear mass $A$ falls below about 30
\citep{Holmes76}. We found that, in practice, this thermal
approximation cannot be used at all for $A < 20$ since far too many
lines are predicted in the simulation.  On the other hand, the
relatively small number of individual levels involved means that data
on experimental cross-sections to individual levels of $A < 20$ nuclei
are in some cases quite complete.  \cite{Ramaty79} reviewed these
cases, and we have obtained from their compilation the distribution of
excited states for the nuclei in Table~\ref{ramaty_isot_table} when
these are produced by spallation. The de-excitation of all other
nuclei with $A < 20$ was omitted from the simulation.

The statistical model is also expected to break down for low incident
particle energies (below a few MeV). In particular, this applies to
spallations that are induced by secondary neutrons and protons. In
these cases individual levels, rather than a statistical distribution
parameterized by $T$, will be excited. In addition, the difference in
level distributions produced by protons and neutrons (which we have
otherwise ignored) will be important.

\subsection{\label{orihet} ORIHET}

The original ORIHET program was developed for the GGOD Monte
Carlo suite \citep{Lei96, Dean03}. It is an adaptation of the Oak
Ridge Isotope Generation and Depletion code ORIGEN, which was designed
to calculate the build-up and decay of activity in any system for
which the nuclide production rates are known. To do so, ORIHET has a
built-in data base containing the half-lives and decay channels of a
large number of radio-isotopes. Currently, ORIHET returns the activity
at a given time for two different radiation histories:
constant irradiation, and ``cooling down'' after constant
irradiation. In the first case, it is assumed that nuclei are produced
at a constant rate for a given time period. 
This model is useful for simulating instrumental backgrounds for
missions in radiation environments that are relatively stable, or for
deriving average instrumental backgrounds for missions exposed to
variable radiation fields as are encountered e.g.\ in LEOs.
In the second case, it is assumed that nuclei are produced at
a constant rate for a given period of time during which production and
decay compete, then the production ends and the radio-isotopes decay
for a given ``cool-down'' period. 
This model is e.g.\ useful for estimating the instrumental background
due to transits through the SAA. 

The original ORIHET radionuclide data base did not include all
isotopes relevant for simulations of instrumental backgrounds in
\g-ray experiments. Specifically, for some isotopes which can be
produced in their ground state as well as an isomeric state, such as
$^{24}$Na, the isomeric state was absent from the original data base.
For MGGPOD, we have changed ORIHET to include all isomers missing in
the original data base which produce significant \g-ray lines
identified in spectra of TGRS or other Ge spectrometers. In addition,
we corrected the treatment of $\alpha$-decays, which in the original
data base were treated as $\beta$-decays.

\subsection{\label{decay} DECAY}

As was ORIHET, the original DECAY package was created for the GGOD suite
\citep{Lei96, Dean03}. The DECAY package is the analog of PROMPT for
simulating radioactive decays. It consists of a data base containing
information on the decays of a large number of radioactive isotopes,
and of code to access these data and to generate random samples of
decay particles. Linking the MGEANT, GCALOR, and DECAY packages into a
single executable (see Fig.~\ref{mggpod_flow_chart} and
Table~\ref{mggpod_sim_steps}) enables one to simulate the decays of a
given radio-isotope with a given activity over a given time period at
any location in the geometric set-up.

To generate the DECAY data base, information concerning the decay
schemes, such as branching ratios for $\beta^-$, $\beta^+$ and
electron capture (EC) decays, internal transition (IT), and
$\alpha$-decay channels, as well as the energy levels and \g-ray to
internal conversion electron branching ratios, was taken from
ENSDF. These data have been supplemented with information on X-ray and
Auger electron fluorescence yields and energies as given in
\citet{Firestone96}.

In developing MGGPOD, we improved and expanded the DECAY package. The
number of isotopes in the data base was increased, and data base and
code upgraded to simulate approximatively isomeric levels (these
sometimes give rise to significant instrumental lines in \g-ray
detectors) as described below.  Information on the emission of
internal conversion electrons is very incomplete in ENSDF data. For a
few isotopes relevant for Ge detectors we corrected the data files
(e.g.\ for the 691~keV E0 transition in $^{72}$Ge, for which photon
emission is not possible).

The de-excitation cascade of a daughter nucleus may proceed through
one or more isomeric levels, which can have a significant effect on
the time sequence of the emission of de-excitation particles. DECAY
has the capability to account for isomeric levels in a de-excitation
cascade in an approximate way by simulating the de-excitation
particles in batches or groups; PROMPT applies the same approach to
prompt de-excitations. In the simulations, a detector time resolution,
$\tau$, can be defined (which approximates the shaping time of the
detector electronics). Levels for which no lifetime information is
available are assumed to have negligible lifetime and hence to decay
instantaneously. For a level with finite lifetime, each time a cascade
is simulated a decay time, $t$, is randomly calculated based on the
lifetime, i.e.\ it is assumed that once the de-excitation cascade
reaches this level the cascade does not proceed until a time $t$ has
passed. The simple algorithm to group the decay particles is as
follows. If no level with finite lifetime is involved, all
de-excitation particles are assumed to be emitted instantaneously.  If
one level with finite lifetime is involved, in case $t \ge \tau$ the
particles are assumed to be emitted in two batches (all particles
emitted in the transitions from the initial level to the isomeric
level, and all particles emitted in the subsequent transitions from
the isomeric level to the ground state of the daughter nucleus);
otherwise ($t < \tau$) all particles are assumed to be emitted within
$\tau$ and hence are started simultaneously in the simulation.  If two
or more levels with finite lifetime are involved, the procedure is
slightly more complicated and becomes iterative. Let us assume there
are $n$ such levels, with randomly selected decay times $t_i, i = 1,
\dots, n$. If $t_1 \ge \tau$, then all particles emitted in the
transitions from the initial level to this first isomeric level are
combined into a first particle batch and started simultaneously in the
simulation. Further particle grouping commences with the first
isomeric level assuming the role of the initial level and $n-1$
remaining potential isomeric levels. If $t_i <
\tau$, a level $m$ is searched such that $\sum_{i=1}^m t_i \ge
\tau$. If such a level exists, all particles emitted in the
transitions from the initial level to isomeric level $m$ are emitted
within $\tau$ and hence started simultaneously in the
simulation. Subsequent grouping commences with isomeric level $m$
assuming the role of the initial level and $n-m$
remaining potential isomeric levels. If no such level exists, i.e.\
$\sum_{i=1}^n t_i < \tau$, all particles are emitted simultaneously.

%%%%%%%%%%%%%%%%%%%%%%%%%%%%%%%%%%%%
%   application to modeling TGRS   %
%%%%%%%%%%%%%%%%%%%%%%%%%%%%%%%%%%%%

\section{\label{mod_tgrs} Modeling Data of the Ge
Spectrometer TGRS}

\subsection{\label{tgrs_description} Instrument and Radiation
Environment}

The Transient Gamma-Ray Spectrometer (TGRS) on board the {\it Wind}
spacecraft was primarily designed to perform high-resolution
spectroscopy of transient \g-ray events, such as \g-ray bursts or
solar flares \citep{Owens95}.  The detector itself consists of a
215~cm$^3$ high purity n-type Ge crystal sensitive to energies in the 
20~keV -- 8~MeV band, kept at its operating temperature of 85~K by a
passive radiative cooler constructed mainly of Be and Mg.  
Some shielding in the soft X-ray range, mainly against intense solar
flare X-rays, is provided by a 30~mil (0.762~mm) thick sheet of Be-Cu
alloy around the sides of the cooler.  The TGRS detector is located on
the south-facing surface of the rotating cylindrical {\it Wind} body,
which points permanently toward the southern ecliptic pole. The
spectrometer has no active shielding and is permanently exposed to
$\sim 1.8\pi$~sr of the southern hemisphere which is unobstructed by
the cooler. A 1~cm thick Pb occulter attached to the {\it Wind} body
exploits the spacecraft rotation in order to modulate the signal from
the ecliptic plane (including the Galactic Center); as seen from the
detector it occults a band $90^{\circ}$ long by $16^{\circ}$ wide
sweeping out the whole ecliptic in {\it Wind}'s 3~s rotation period.
The chemical elements mentioned, together with Al which was used for
most structural components, are expected to be among the main sources
of prompt and delayed \g-ray line backgrounds.

Since its launch on November 1, 1994, {\it Wind} has been following
unusual and highly elliptical orbits;
halo orbits around the Earth-Sun Lagrangian L1 point in the first half
of the mission, trans-lunar Earth orbits in the second half
\citep{Acuna95}. Hence {\it Wind} spent virtually the whole mission in
interplanetary space, well away from near-Earth radiation backgrounds
such as geomagnetically trapped particles and Earth albedo radiations.
The radiation environment experienced by TGRS therefore is dominated
by two components: diffuse cosmic hard X and \g\ radiation, and
Galactic cosmic rays, unmodulated by the Earth's magnetic field. Both
radiation fields can be considered isotropic. The diffuse cosmic hard
X and \g\ radiation is constant in time. Significant secular changes of
Galactic cosmic rays due to solar modulation occur on time scales of a
few months.  These changes were observed by TGRS in several \g-ray
lines \citep{Harris01}. There have been very few interruptions in the
data stream, generally caused by brief passages of the trapped
radiation belts around perigee or by triggering of a special mode of
data collection by solar flares and \g-ray bursts.

The background spectra recorded by TGRS provide an ideal test for the
MGGPOD package. The (relative) simplicity of the instrument design
facilitated the development of an accurate model for Monte Carlo
simulations (this also applies to the {\it Wind} spacecraft). The
stability and isotropy of the radiation environment, which can be
represented by only two dominant components, also greatly simplify
Monte Carlo simulations, and at the same time render quantitative
comparisons of TGRS spectra with simulations more rigorous. Finally,
the fine energy resolution of the detector reveals the plethora of
instrumental lines in great detail, providing us with sensitive tests
on the numerous interaction channels through which Galactic cosmic
rays can deposit their energy in the instrument and spacecraft
structures. The resolution of TGRS at 500~keV was nominally about
3~keV full width at half maximum (FWHM), which was achieved in the
early months of the mission \citep{Harris98}; thereafter resolution
degraded due to accumulated damage from cosmic-ray impacts
\citep{Kurczynski99}. The line profiles also became distorted, with
marked tails on the low-energy wings, and energy calibration became
more difficult due to gain shifts.  We limit our analysis to the
period Jan.--May, 1995, when these problems were negligible,
except for small low-energy tails particularly on the highest energy
lines (see \S~\ref{tgrs_line-ids}). 
The TGRS background spectra were binned in 1~keV energy channels over
the entire 20~keV -- 8~MeV range. However, due to problems with
saturation of electronics components after very high energy deposits
by heavy Galactic cosmic rays the energy range of TGRS for scientific
studies had to be limited to 40~keV -- 8~MeV. In addition, the
210--260~keV range is contaminated by electronic artifacts, which
appear as broad features in the spectrum. Our reference Jan.--May,
1995 TGRS spectrum is shown in Fig.~\ref{compare_tgrs_sim_spc}.

The detector cannot distinguish individual energy deposits which occur
within the peaking time of the electronics, which is about
$5~\mu$s. In general, nuclear transitions within the Ge crystal from
excited states to the ground state will be detected at the sum of the
energies if they proceed through one or more intermediate levels. An
exception occurs for transitions via states with lifetimes $\tau \ga 5
~\mu$s, which may survive to be detected separately; we refer to these
states as isomers for the purpose of working with TGRS. This is
particularly relevant for the de-excitation of $^{73m}$Ge: the
de-excitation of the 66.7~keV level ($\tau = 0.72$~s) proceeds via a
level at 13.3~keV with $\tau = 4.3~\mu$s. If the 13.3~keV level takes
long to de-excite, the decay will be registered as two distinct
events, giving rise to lines at 13.3~keV and 53.4~keV; if it
de-excites quickly the two photons will be summed giving rise to a
single line at 66.7~keV; for intermediate cases the energies are only
partially summed and the event will fill in the interpeak region
between 53.4~keV and 66.7~keV.  Radioactive $\beta^-$ or $\beta^+$
decays within the detector do not give rise to line features in the
spectrum; the energy of the $\beta$ particle is distributed
continuously and summed with the coincident \g\ rays of discrete
energy.  Electron capture $\beta$ decays within the Ge crystal do
produce multiple line features.  Electron captures are accompanied by
X-rays and/or Auger electrons arising from the filling of a vacancy in
one of the atomic sub-shells; lines will appear at the sum of the
nuclear transition energy and the binding energy of the atomic
sub-shells. Certain lines will appear in the TGRS spectrum which are
not due to cosmic-ray effects, notably a strong line at 1460~keV from
a $^{40}$K source which was flown for calibration purposes, and lines
from the decay chains of U and Th occurring naturally in traces, e.g.\
in the Be cooler and shield.

\subsection{\label{tgrs_sim} Models and Simulation}

In modeling the TGRS background by Monte Carlo simulation
approximations regarding the mass model representing the instrument
and the {\it Wind} spacecraft, as well as the components of the
radiation environment and their time history, are inevitable.  All
elements of the TGRS instrument described above and of the {\it Wind}
spacecraft are sources of prompt and delayed instrumental
backgrounds. The required mass model has to give a faithful
representation of the geometrical structure and its atomic/isotopic
compositon, particularly for the Ge detector and its vicinity. With
increasing distance from the detector the details of the mass
distribution are less critical; however, we were careful to conserve
the total mass of the spacecraft components and to represent their
atomic/isotopic composition, as these are important factors for the
generation of secondary particles and residual nuclei in hadronic
interactions. Our model of the TGRS detector is based on the mass
model created by \citet{Seifert95} for response matrix
generation. This early mass model replicated well the geometrical
structure and the masses of the TGRS instrument above and around the
Ge crystal as described by \citet{Owens95}. However, it did not
specify the atomic and isotopic compositions of materials in
sufficient detail for our purpose, it did not include instrument
components below the Ge crystal, such as electronics boxes, and it
only provided the crudest representation of the spacecraft. We
supplemented and improved the early TGRS mass model by referring to
original technical drawings and other documentation, and by close
examination of spare parts.  Assuming that the latest documents on the
detector mass that were available to us are the most accurate, we can
account for about 96\% of the mass of the total instrument (18.1~kg
out of 18.9~kg). However, in and close to the detector, i.e.\ the
areas which are most important for the simulation, we can account for
only 86\% of the material (6~kg out of 7~kg). Our approach to creating
a model for the {\it Wind} spacecraft and other scientific instruments
was similar. For example, the basic geometry and structure of the
spacecraft body are described in
\citet{Harten_Clark95}. We can account for about 86\% of the
documented spacecraft mass
in our mass model.

The radiation environment of TGRS was approximated by its two dominant
components: diffuse cosmic hard X and \g\ radiation, and cosmic
radiation. Both were assumed to be isotropic and constant in time. We
used an analytic approximation by \citet{Gruber99} to model the
spectrum of the diffuse cosmic photon radiation in the 30~keV to 9~MeV
energy range.  Due to limitations of GCALOR (see \S~\ref{gcalor}) the
only hadronic particle component of the cosmic radiation that could be
simulated were protons (the contribution of cosmic-ray electrons to
TGRS spectra was found to be negligible). We used a proton spectrum
(energy range 10~MeV to 226~GeV) that was calculated for early 1995
(Moskalenko 2002, priv.\ comm.) with the GALPROP Galactic cosmic-ray
progagation code, taking into account solar modulation using a
steady-state drift model \citep{Moskalenko02}. The GALPROP code has
been shown to reproduce simultaneously observable data of many kinds
related to cosmic-ray origin and propagation
\citep{Moskalenko_Strong00}. This model proton spectrum was
assumed to be representative of the Jan.--May, 1995 time period, during
which the TGRS data we are modeling were recorded. During this time
period the activity of some isotopes produced by cosmic radiation
built up, and hence their contribution to the TGRS data increased with
time. However, the TGRS spectrum we are modeling represents an average
over this time period. In our simulations of the instrumental
background due to delayed radioactive decays we approximate this time
average with a snapshot in March~1995 after four months of cosmic-ray
irradiation, i.e.\ we use isotope activities that are calculated
assuming four months of constant irradiation with our model cosmic-ray
proton spectrum.

\subsection{\label{tgrs_line-ids} Identifications of TGRS Instrumental 
Lines}

One of the main goals for developing the MGGPOD suite was to create a
Monte Carlo tool capable of modeling the many instrumental lines
present in \g-ray detectors. To assess the extent to which this goal
was reached for TGRS we attempted to identify its more than 200
observable lines and spectral features. The TGRS spectra were analyzed
using the GASPAN\footnote{The software, developed by F.~Riess, and
documentation are available under {\tt http://ftp.leo.org/download/
pub/science/physics/software/gaspan/}; alternatively, F.~Riess can be
contacted directly (friedrich@die-riessens.de)} gamma spectrum
analysis program. GASPAN allowed us to characterize the shape of the
instrumental lines as a function of energy, following parametrizations
given by \citet{Phillips_Marlow76}. The photopeak component of each
line was described by a Gaussian and a low-energy ``wing'', whose
importance increased with energy. The wing was modeled by the sum of
two exponential tails representing effects from multiple Compton
scatterings and incomplete charge collection. The line width and tail
parameters were determined as a function of energy using the strongest
and cleanest lines. This energy dependent line shape was then employed
to analyse TGRS spectra, particularly to resolve the many closely
spaced or blended lines.  When analyzing intrinsically broad lines
arising from the prompt de-excitation of very short-lived levels
\citep[$T_{1/2} \la 0.5$~ps,][]{Evans02} allowance had to be made for
their exceptional width, which is caused by Doppler broadening due to
the motion of the nucleus during de-excitation.

The results of our spectral analysis are summarized in
Tables~\ref{class1_ids} and \ref{class2_ids}, in which the lines are
identified and the count rates in the observed spectrum are presented.
Lines or blended features for which the GASPAN fit determined a
significance $< 5 \sigma$ are not included.  Most of the strong lines
had already been identified with nuclear transitions in previous work
with Ge spectrometers \citep{Wheaton89, Bartlett94, Evans02,
Weidenspointner03}, with which our results are in general agreement.
However, the level of detail with which our simulation was performed
allowed us to suggest that in many cases multiple transitions can be
identified contributing to previously unidentified blends.  Thus in
Tables~\ref{class1_ids} and
\ref{class2_ids} two or more transitions are frequently assigned to an
observed line feature.  The order in which the transitions are listed
corresponds to their ranking as contributors to the line strength,
according to the simulation,
with transitions whose contribution was $<10$\% according to the
simulation being omitted. Table~\ref{inelast_recoil_table} lists
identified triangular or ``sawtooth'' shaped features from
inelastic neutron scattering off Ge nuclei in the detector crystal, as
described in \S~\ref{prompt_ncap_insct}.

Narrow (unresolved) lines were assigned to Table~\ref{class1_ids} if
one or more identifications could be made, from earlier work or from
the simulation.  If no certain identification could be made they were
put in Table~\ref{class2_ids}, whose purpose is to draw attention to
gaps in our knowledge of nuclear physics to encourage further
work. Lines which were visibly broader than the TGRS energy resolution
were treated as blends, keeping in mind that an intrinsically broad
component might also contribute.  If multiple identifications were
possible, sufficient (in theory) to explain the line width, the line
was included in Table~\ref{class1_ids}. Two indications were used to
assign a blend to the ``uncertain identity'' category
(Table~\ref{class2_ids}).  First, a blend which was obviously broad
enough to contain multiple lines might correspond to only one line in
the simulation (or none).  Second, a comparison of the observed and
predicted line strengths might show the simulation seriously failing
to reproduce an observed line.  Although there is considerable scatter
in the reliability of the simulations even for well-identified narrow
lines (see below), we somewhat arbitrarily assigned blends where the
ratio of simulated to observed line strength fell below 25\% to
Table~\ref{class2_ids}.

There are some obvious deficiencies in this qualitative procedure by
which Tables~\ref{class1_ids} and \ref{class2_ids} were compiled.
There are considerable quantitative uncertainties in the simulations,
as we shall see, preventing the ranking of transitions and the 10\%
cutoff from being totally reliable.  The simulation may even omit
lines altogether.  Such cases belong in Table~\ref{class2_ids}, but
may appear in Table~\ref{class1_ids} if the line appears narrow or if
$>25$\% of its observed strength is predicted by the simulations.
Several cases of this kind may be indicated by measurements of the
line energy in Table~\ref{class1_ids} which are inconsistent with the
transition energies.  However, these cases may also be due to problems
inherent in the use of the rather complicated function
\citep{Phillips_Marlow76} for fitting the line profiles.

\subsection{\label{tgrs_sim-vs-data} Comparison of Simulation and
Data}

%A comparison of the Jan.--May~1995 TGRS spectrum with a MGGPOD
%simulation is shown in Fig.~\ref{compare_tgrs_sim_spc}. 
A comparison of the overall Jan.--May, 1995 TGRS spectrum with the
various components of our MGGPOD simulation is shown in
Fig.~\ref{compare_tgrs_sim_spc}. The simulated instrumental background
components are: prompt background due to cosmic-ray proton
interactions and prompt de-excitations (green), delayed backgrounds
from radioactive decays in the TGRS instrument and the {\it Wind}
spacecraft (purple), and background due to diffuse cosmic X and \g\
rays (blue). The sum of all components is depicted in red, and the data
are shown in black (the broad features in the 210--260~keV region are
electronic artifacts). For better comparison, a model of the
instrumental resolution has been applied to the simulation
results. This model includes the anomalous width of the 511~keV line,
which arises from the finite momenta of positron and electron at
annihilation, but does not include Doppler broadening of lines due the
motion of the nucleus during de-excitation from short-lived levels
with $T_{1/2} \la 0.5$~ps \citep{Evans02}, as is the case e.g.\ for
the 4438~keV line from $^{12}$C$^\ast$. Lines due to radio-isotopes of
the so-called natural decay chains (e.g.\ $^{212}$Pb) and due to the
$^{40}$K calibration source are not included in the simulations.
\notetoeditor{Fig. 5 should be placed at the top of a page, with a
width equal to the width of the text on the page.}
Comparisons of the Jan.--May, 1995 TGRS spectrum in smaller energy
ranges with the summed MGGPOD simulation are provided in
Figs.~\ref{compare_tgrs_sim_spc_40-800}--\ref{compare_tgrs_sim_spc_3200-8000}.
In these figures, some of the more prominent lines and spectral
features have been labelled for easier reference to the detailed
identifications in Tables~\ref{class1_ids} and \ref{class2_ids}.

\subsubsection{Continuum and Inelastic Neutron Scattering Features}

The MGGPOD simulation reproduces very well the overall shape and
magnitude of the actual background.  This is illustrated in
Fig.~\ref{compare_tgrs_sim_spc_ratio}, which depicts the ratio between
the simulation and the measured spectrum. Disregarding line features,
this ratio exhibits very little trend with energy and has a value of
about 0.85. The only exception occurs at the lowest energies, below
about 200~keV, where the ratio increases with decreasing energy up to
a value of 1.2.

Below about 200~keV diffuse cosmic photons are found to be the
dominant source of instrumental background in TGRS, as expected for a
wide field-of-view instrument. Prompt cosmic-ray proton induced
events are the main background component at energies above about
200~keV, and practically the sole background above about 4~MeV. The
relative dominance of the prompt background in TGRS reflects the lack
of any veto shielding. The prompt background comprises different
processes, among them direct ionization in the detector by cosmic-ray
protons and energy losses by secondaries, including energy deposits
due to elastic and inelastic scattering of secondary neutrons off Ge
nuclei in the detector crystal. The prompt background is not a pure
continuum, but features numerous lines; these will be addressed in
\S~\ref{tgrs_sim-vs-data_lines}.

As discussed in \S~\ref{prompt_ncap_insct}, inelastic neutron
scattering gives rise to characteristic ``saw-tooth'' or
triangular-shaped broad features in the spectrum. Their absolute and
relative strengths, and less so their detailed shape, are sensitive to
the flux and spectral distribution of secondary neutrons in the Ge
crystal at energies around 1~MeV. In Table~\ref{inelast_recoil_table}
we attempt to make comparisons between our measurements of the
triangular inelastic neutron scattering features and the simulations.
Fig.~\ref{compare_tgrs_sim_spc}, and in more detail
Figs.~\ref{compare_tgrs_sim_spc_40-800}--\ref{compare_tgrs_sim_spc_1100-1900},
make it clear that upon these 
features there are superposed a large number of strong lines, that
underlying them there must be a continuum of somewhat uncertain shape,
and that the features sometimes overlie each other.  Nevertheless,
there is sufficient resemblance between the simulated and observed
recoil features for comparisons to be made -- lower and higher energy
bounds and lines can be consistently identified, hence simulation and
observation can be treated consistently.

Our method involved simply identifying the lower and upper energy
bounds of the recoil features, summing the counts between these
limits, and subtracting the estimated counts contributed by the lines
and the continuum.  The uncertainties in the values in
Table~\ref{inelast_recoil_table} are overwhelmingly dominated by the
uncertainties in the true values of line and continuum count rates.  We
quote ranges of count rate in each recoil feature corresponding to the
most extreme estimates of the line and continuum strengths.  The
extreme low continuum estimate assumes a flat continuum having the
flux corresponding to the upper energy bound; the high continuum
estimate is a linear interpolation between the lower and higher energy
bounds.  The line measurements within the recoil feature energy ranges
(Table~\ref{class1_ids}) were assumed to be upper limits because of
the possibility of contamination by the recoil feature.  Of course,
upper line and continuum limits correspond to lower limits on the
strength of the recoil feature in Table~\ref{inelast_recoil_table},
and {\em vice versa}.

In Table~\ref{inelast_recoil_table}, note that the upper and lower
energy bounds are somewhat arbitrarily chosen from one feature to the
next, due to the interference of strong lines and other recoil
features. Thus the results should not be compared from one feature to
the next -- the important point is that theory and observation can be
compared within the same feature, since they are treated consistently.

We are left with the conclusion, based on the figures in
Table~\ref{inelast_recoil_table} (where comparable) that the
simulation and the measurement in general agree within 30\% -- an
exception is the one above 834~keV, which suffers exceptionally strong
interference from the lines at 844 and 847~keV (see
Fig.~\ref{compare_tgrs_sim_spc_800-1100}). This estimate is supported
by the lack of any discontinuities in the ratio of simulation and
actual data at the recoil features' energy ranges in
Fig.~\ref{compare_tgrs_sim_spc_ratio}. There is a weak tendency for
the theory to over-predict the strength of the recoil features. This
might result from systematic errors in any of the various quantities
that are relevant for the simulation of the recoil features, such as
the flux of secondary neutrons -- which depends e.g.\ on the accuracy
of the mass model and of hadronic cross-sections in general, and the
Ge neutron cross-sections in particular.

Delayed radioactive decays in the TGRS instrument and the {\it Wind}
spacecraft also give rise to continuum background (\g-ray lines from
radioactive decays are discussed below). The dominant contributors are
$\beta$ decays in the Ge crystal. As described in
\S~\ref{tgrs_description}, $\beta^-$ or $\beta^+$ decays within the Ge
detector result in a pure continuum background (except if isomeric
levels are involved in the de-excitation of the daughter nucleus)
because the continuously distributed energy of the $\beta$ particle is
summed with the coincident \g\ rays of discrete energy.  In the case
of TGRS and {\it Wind}, continuum background due to radioactive decays
in the spacecraft is relatively less important because $m/r^2$, with
$m$ being a mass element and $r$ the distance from the detector, is
lower for the spacecraft structure than for instrument parts. The
delayed background exceeds the diffuse cosmic photon background above
about 400~keV, and cuts off at about 4~MeV.

\subsubsection{\label{tgrs_sim-vs-data_lines} Lines}

The simulation is also very successful in modeling the more than 200
lines that are observed in the TGRS spectrum. Most (about 87\%) of the
lines are reproduced, with the ratio of simulated and actual line
count rates clustering around a value of 1 with no trend in energy.
The simulation produces a few spurious lines, and fails to reproduce a
few lines that are present in the data. 

A quantitative comparison of simulated and observed line strengths is
shown in Fig.~\ref{ratio_simdat_scat_plot}, in the form of a scatter
plot of the ratio between them as a function of line energy. The
distribution of the numerical values of these ratios are shown in
Fig.~\ref{ratio_simdat_hist_plot}.  Overall, the observed lines are
reproduced within a factor $\sim 2.5$.  There is no trend with energy
in Fig.~\ref{ratio_simdat_scat_plot}, and the mean ratio is very close
to 1, suggesting that our methods of line simulation (\S
\S~\ref{decay}, \ref{prompt}) are sound to a first level of
approximation.  There are enough lines (205) for the results to be
broken down into subsets in search of interesting systematic
deviations from these overall conclusions.

If the simulated-to-observed line strength ratios are broken down by
reaction type it is found that lines from radioactive decays are on
average well reproduced by the simulation, while spallation reaction
line strengths are somewhat overestimated
(Fig.~\ref{ratio_simdat_hist_plot}, dotted line).  Leaving aside
possible problems of isotope production from the incoming channel
cross-sections in GCALOR, this effect results solely from a systematic
overestimate of the strengths of lines from low-mass ($20
\le A \le 30$) nuclei, by a factor of nearly 2
(Fig.~\ref{ratio_simdat_hist_plot}, dashed line).
This indicates that the thermal approximation for spallation product
de-excitations (\S~\ref{prompt}) becomes increasingly inadequate with
decreasing mass number below $A = 30$, as might be expected from
nuclear level statistics.  Spallation reactions overall do not show a
wider spread than the others (a factor $\sim 2.5$ in
Fig.~\ref{ratio_simdat_hist_plot}), nor do they show any trend with
line energy.  This indicates that, to the extent that the thermal
approximation is valid, the choice of temperature 3~MeV is justified.

%%%%%%%%%%%%%%%%%%%%%%%%%%%%%%%
%   summary and conclusions   %
%%%%%%%%%%%%%%%%%%%%%%%%%%%%%%%

\section{\label{sum-conc} Summary and Conclusion}

We have described the capabilites, functioning, and structure of the
MGGPOD Monte Carlo simulation suite, which is based on the widely used
GEANT Detector Description and Simulation Tool
\citep[Version~3.21,][]{Brun95}. MGGPOD is a user-friendly simulation
package that is applicable to all stages of space-based
\g-ray astronomy missions. In particular, the MGGPOD package allows
{\sl ab initio} simulations of instrumental backgrounds, both
continuum and \g-ray lines, that arise from interactions of the
various radiation fields within the instrument and spacecraft
materials. It is possible to simulate both prompt instrumental
backgrounds, such as energy losses of cosmic-ray particles and their
secondaries, as well as delayed instrumental backgrounds, which are
due to the decay of radioactive isotopes produced in nuclear
interactions. The MGGPOD package and documentation are available to
the public for download at the Centre d'Etude Spatiale des
Rayonnements ({\tt http://sigma-2.cesr.fr/spi/MGGPOD/}).

To demonstrate its capabilities, we employed the MGGPOD suite for
modeling high resolution \g-ray spectra recorded by the Transient
Gamma-Ray Spectrometer (TGRS) on board {\it Wind} during 1995. We
found that both the continuum and the \g-ray line background are very
well reproduced. Regarding the continuum background, the ratio between
simulation and flight data is about 0.85, except for the lowest
energies below about 200~keV, where the ratio increases with
decreasing energy up to a value of 1.2. The simulation even reproduces
well asymmetric, ``sawtooth'' or triangular-shaped broad spectral
features that are due to inelastic neutron scattering off Ge nuclei in
the detector. Regarding background lines, the simulation reproduces
about 87\% of the observed background lines in TGRS, with the ratio of
simulated and actual line count rates clustering around a value of 1
with no trend in energy. Our results for continuum and lines are as
good as or better than those obtained by \citet{Dyer94} for OSSE, by
\citet{Dyer98} for {\it Mars Observer}, or \citet{Graham97} for {\it
HEAO-3} using other Monte Carlo tools.

Below about 200~keV, diffuse cosmic photons are found to be the
dominant source of instrumental background in TGRS. The simulation
overestimates this component, with the largest deviation of 20\%
occurring at the lowest energies. We attribute the overestimate mainly
to uncertainties in the mass distribution around the Ge
detectors. Although we made a substantial effort to describe the
detector as accurately as possible in our mass model, we may have missed
material in and close to the detector (as described in
\S~\ref{tgrs_sim}). Small changes in the amount and distribution of
mass close to the detector have a large effect on the simulated count
rate at a few tens of keV because of the strong photoelectric
absorption at these energies, and also affect the probability of
Compton scattering of higher-energy photons into the detector.  As
mentioned in \S~\ref{tgrs_sim}, we employed in our simulations an
analytical model of \citet{Gruber99} for the spectrum of the diffuse
cosmic photons, which was obtained by fitting results from various
instruments. The uncertainty of this model is expected to be smaller
than our overestimate of the background due to diffuse cosmic photons;
hence the model is unlikely to be the sole cause of it.

The shortfall of the simulation above 200~keV is expected. Currently,
GCALOR is not capable of simulating hadronic interactions of $\alpha$
particles, hence the contributions of cosmic-ray $\alpha$ particles to
the prompt and the delayed background could not be modeled.  A rough
estimate of this contribution can be obtained with an empirical
formula for the ratio of the total $\alpha$ to proton cross-sections
by \citet{Ferrando88}. This ratio does not vary much with the mass of
the target nucleus between Al and Ge, and has an average value of
about 1.7. By number, the ratio of cosmic-ray $\alpha$s to protons is
about 0.14; hence we can estimate that if cosmic-ray $\alpha$
particles could be included in the simulation, they would increase the
prompt and delayed background by roughly 24\% -- enough to account for
the current short-fall of about 15\%. The effect of $\alpha$ induced
spallations on the activation of specific isotope species is very
difficult to estimate, since the ratio of $\alpha$ to proton
cross-sections for specific spallation products strongly varies with
energy \citep{Ferrando88}.  Similarly, when modeling data from the
OSSE instrument on board {\it CGRO},
\citet{Dyer94} concluded that $\alpha$ particles account for about
20\% of the calculated activation due to cosmic rays.  In addition,
there are deficiencies in the mass model, since the total mass
accounted for in the instrument and the spacecraft falls short of the
documented mass budget by about 14\% (\S~\ref{tgrs_sim}). Less mass
will result in fewer interactions and hence less prompt background as
well as less activation and therefore less delayed background. Except
for the lowest energies, shielding of the detector from \g-rays
originating in different parts of the detector and spacecraft is not
expected to be significantly affected by the moderate mass deficit.

The simulation reproduces about 87\% of the observed
background lines in TGRS, with the ratio of simulated to measured line
count rates clustering around a value of 1 with no trend in
energy. On average, lines from radioactive decays are well
reproduced. We see a small trend of overproduction for lines from
de-excitation of spallation products with decreasing mass. This trend
is expected given the simple statistical method that is used for
generating de-excitation photons. 

The simulation of lines from de-excitations involving isomeric levels
is influenced by the choice for the detector time constant
(\S~\ref{tgrs_description}). In our simulations, we set this parameter
to $\tau = 5~\mu$s; hence the results for lines from the isomeric
levels of $^{73}$Ge (the 13.3~keV level has a lifetime of $4.3~\mu$s)
are particularly sensitive to this parameter. As can be seen in
Figs.~\ref{compare_tgrs_sim_spc}, \ref{compare_tgrs_sim_spc_40-800},
and \ref{compare_tgrs_sim_spc_ratio}, both lines at 53.4~keV and
66.7~keV are overestimated, with the sum peak being relatively too
strong. In the simulation, partial summation of energy deposits by the
electronics, which fills the interpeak region, is not modeled, thereby
overestimating the sum peak. However, to reproduce the shape of the
$^{73m}$Ge complex better in the simulation, the parameter $\tau$
needs to be set to a value smaller than $5~\mu$s.

There are a few lines which allow us to assess the overall quality of the
simulation. The 511~keV line, which results from a multitude of
processes (e.g.\ decay of $\pi^+$ produced in hadronic interactions,
electromagnetic showers, radioactive decays) is reproduced within a
factor of 0.74, indicating that there are no severe problems with
modeling the diverse physics involved. 
The 2.223~MeV line from thermal neutron capture on H, the features
from inelastic neutron scattering off Ge, and the Doppler broadened
line at 2.211~MeV line from inelastic neutron scattering on $^{27}$Al
are reproduced within factors of 1.4, $\la 1.3$, and 1.1,
respectively. This indicates that flux and energy spectrum of
secondary neutrons throughout TGRS and {\it Wind} are fairly well
reproduced, which in turn implies that there are no severe problems
with the treatment of neutron production, propagation, and
interaction.

The remarkable performance of MGGPOD in modeling TGRS data encouraged
us and others to employ this Monte Carlo suite to other
instruments. Preliminary results for the spectrometer SPI on board
{\it INTEGRAL} have been reported by
\citet{Weidenspointner04a}. MGGPOD reproduces well early-mission SPI
data; however, remaining discrepancies imply that in the case of this
much heavier and complex mission (compared to TGRS) there
are deficiencies in the production and/or thermalization and capture
of secondary neutrons. This has lead us to an ongoing systematic
evaluation of the neutron cross-sections in GCALOR, whose results will
be presented in a future publication of a re-visited SPI background
simulation. Recently, MGGPOD has also been applied to {\it RHESSI} by
\citet{Wunderer04}. 
These preliminary results are encouraging, the overall spectrum is
well reproduced. As expected for a mission in low-Earth orbit
\citep[e.g.][]{Weidenspointner01}, the background is dominated by
decays of radioactive isotopes produced during SAA passages. However,
both SAA induced continuum and line components are overproduced,
indicating the need for refining the modeling of the time-dependent
activation during SAA passages.  Two of us (G.\ W.\ and M.\ J.\ H.)
will use MGGPOD to estimate instrumental background and sensitivity of
the space based gamma-ray lens MAX \citep{vonBallmoos04}.  The
performance and user-friendliness of MGGPOD also attracted the various
instrument teams studying concepts for an advanced Compton telescope;
they selected MGGPOD as their baseline simulation tool for predicting
the performance and sensitivity of the various designs. To address
better the requirements for studying the performance of these complex
instruments, Version 1.1 of MGGPOD, which is currently under
development, will provide additional beam geometry and spectral
options, will include the capability of modeling polarized photons,
and will support additional output formats suitable for the advanced
event reconstruction algorithms \citep[e.g.][]{Zoglauer04} that are
foreseen for these instruments.

%%%%%%%%%%%%%%%%%%%%%%%%
%   acknowledgements   %
%%%%%%%%%%%%%%%%%%%%%%%%

\acknowledgements

We are grateful to David Palmer, Helmut Seifert, and Juan Naya for
helpful discussions about the TGRS instrument. We are indepted to
Prof.\ Friedrich Riess for enhancing GASPAN according to our needs, and
for user support.

\clearpage

%% No more than seven \figcaption commands are allowed per page,
%% so if you have more than seven captions, insert a \clearpage

%% There must be a \figcaption command for each legend. Key the text of the
%% legend and the optional \label in curly braces. If you wish, you may
%% include the name of the corresponding figure file in square brackets.
%% The label is for identification purposes only. It will not insert the
%% figures themselves into the document.
%% If you want to include your art in the paper, use \plotone.
%% Refer to the on-line documentation for details.

%--- MGGPOD flow chart -----------------------------------------------

\figcaption[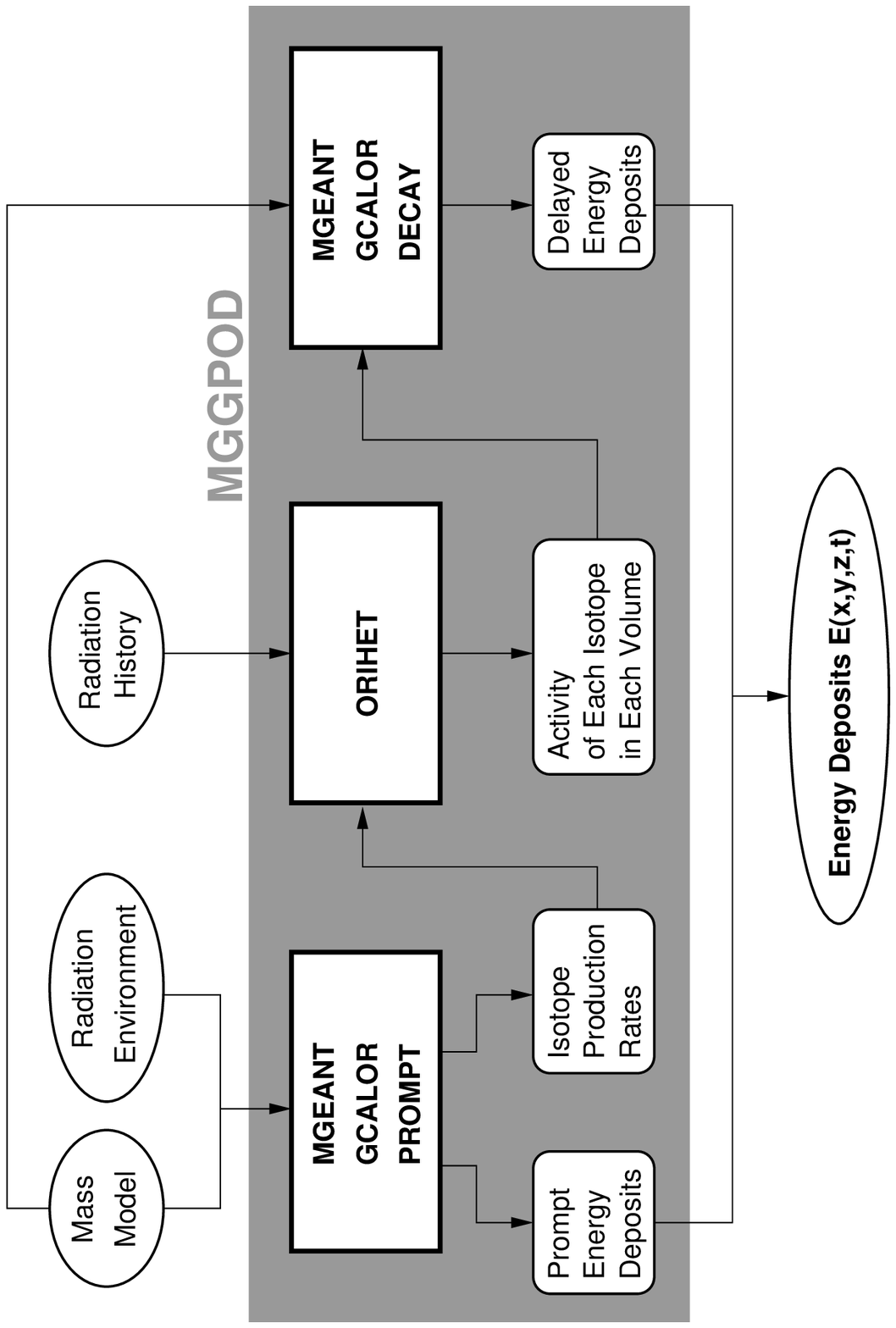]{A flow chart illustrating the overall
structure of the MGGPOD Monte Carlo simulation suite. The various
simulation packages (shown in boxes) and input and output files (shown
in ellipses and round-edged boxes) are explained in the text.
\label{mggpod_flow_chart}}

%--- neutron capture scheme ------------------------------------------

\figcaption[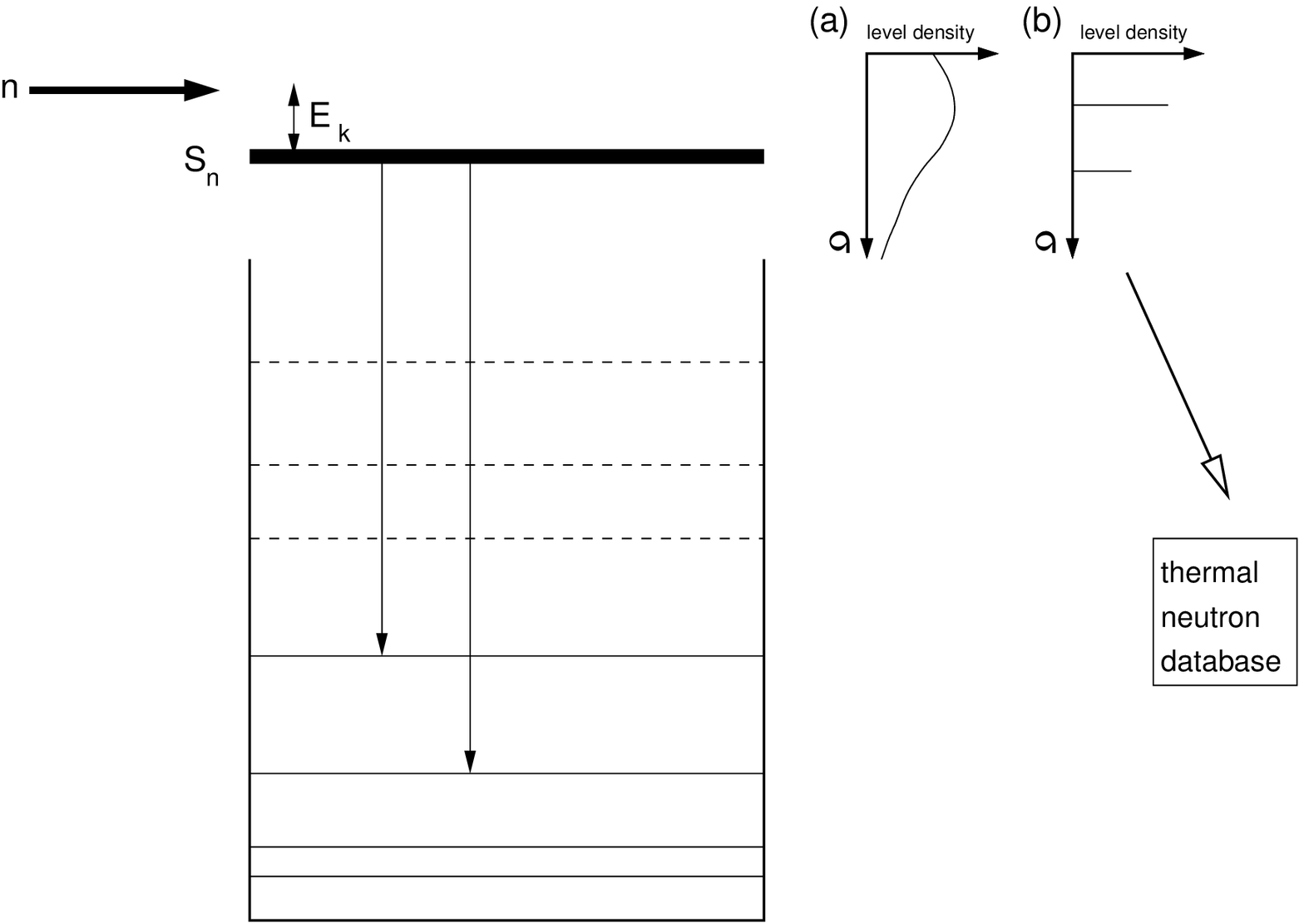]{Schematic representation of our treatment of
(n,$\gamma$) reactions.  The incoming neutron (left, bold arrow) is
captured in a level at excitation energy $S_n + E_k$ in the compound
nucleus, where $S_n$ is the neutron separation energy and $E_k$ the
neutron's kinetic energy. Spin $\sigma$ and parity $\pi$ are assigned
to this level using the level density formula of
\citet{Mughabghab_Dunford98} to obtain a random value (inset (a)).  In
the general case the level is then allowed to decay
electromagnetically to known levels of the nucleus (arrows down) which
have compatible spin and parity according to the electric dipole
selection rules; the probability of transition to any level is
proportional to $E_\gamma^3$.  In the special case (b), where the
$J^\pi$ of the excited level are compatible with s-wave neutron
capture, the downward transition probabilities are obtained directly
from measured thermal neutron capture data (ENSDF).
\label{ncap_scheme}}

%--- inelastic neutron scattering scheme -----------------------------

\figcaption[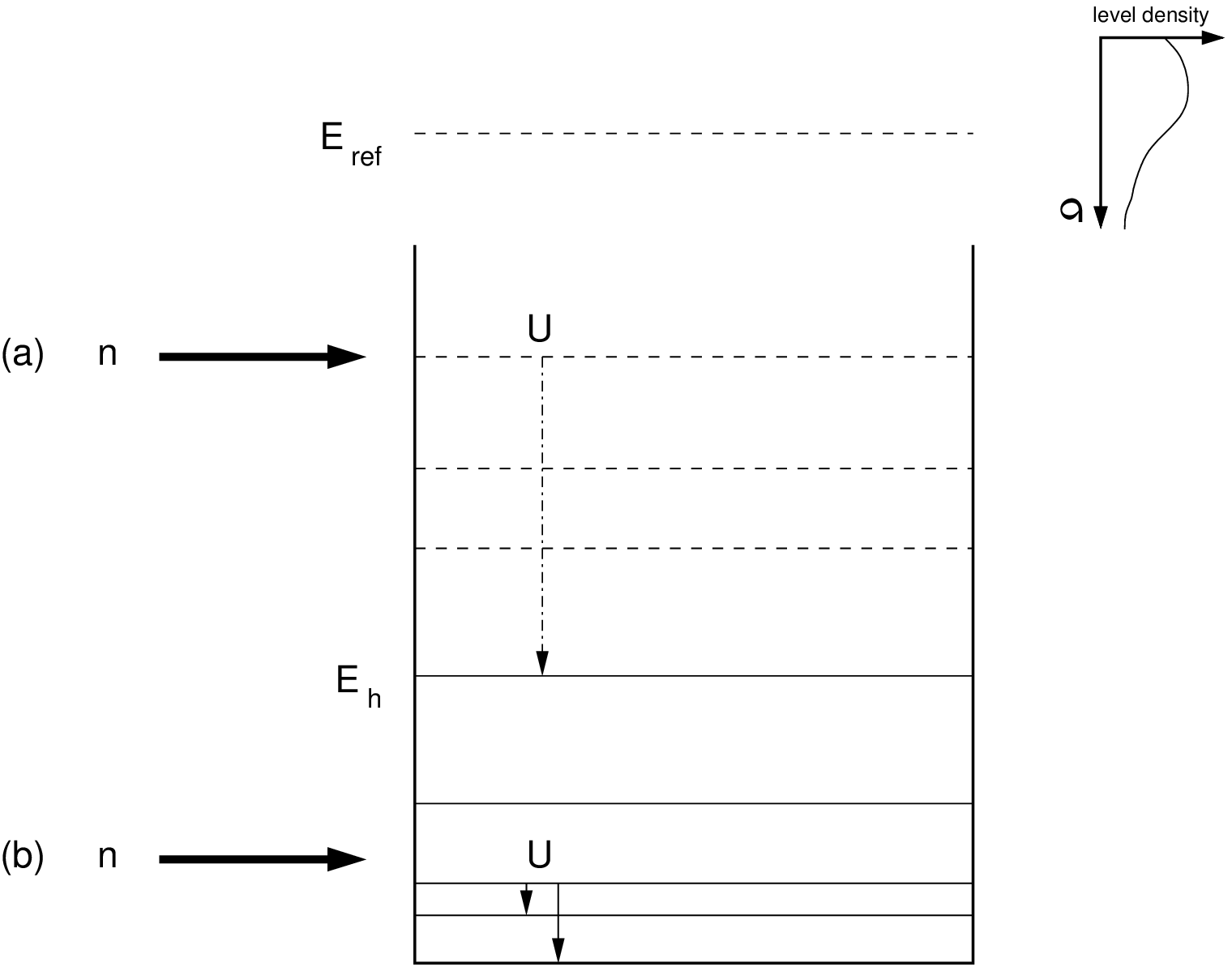]{Schematic representation of our treatment of
neutron inelastic scattering.  The excitation produced in the target
nucleus is known from the GCALOR particle tracking.  Two cases arise.
(a) The incoming neutron (left, bold arrow) excites a level in the
target which lies above the highest known level $E_h$.  The excited
level is assigned a spin and parity at random from the level density
distribution (curve inset, top right).  The level density formula is
evaluated at energy $E_{\mathit{ref}} \sim 10$~MeV, approximately where the
inelastic cross-section peaks.  A de-excitation transition is assigned
to the highest known level with $J^\pi$ compatible with an E1
transition (dashed arrow).  (b) The incoming neutron deposits energy
of a known level (or between known levels); the nearest level in
energy is assumed to be excited.  In either case, once a known level
has been occupied the cascade of transitions down to the ground state
is taken from the ENSDF database.
\label{insct_scheme}}

%--- spallation scheme -----------------------------------------------

\figcaption[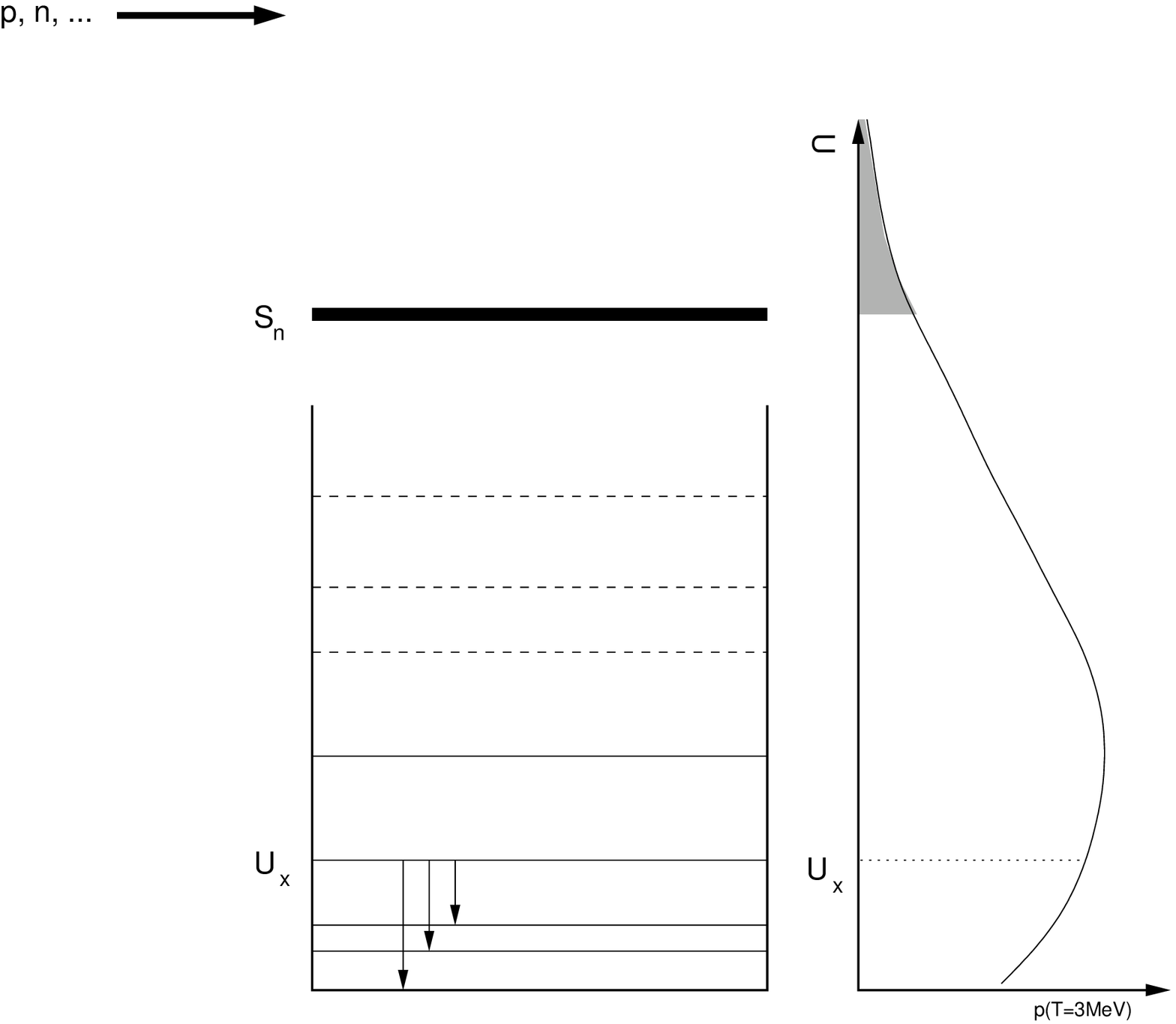]{Schematic representation of our treatment of
spallation reactions.  The incoming particle (left, bold arrow) at
very high energies leaves the compound nucleus in a thermal
distribution of excited states which loses energy by particle
emission.  When it has lost enough energy that only a small "tail"
remains above the neutron separation energy $S_n$, particle emission
ceases.  The thermal distribution function shown at right (rotated)
then applies (the "tail" being shaded), and the known levels (full) in
the remnant nucleus are populated according to a temperature $T =
3$~MeV.  De-excitations from these known levels (an example at energy
$U_x$ is shown) are then taken from the ENSDF nuclear database. We do
not consider the unknown levels (dash) because there are no data for
them in ENSDF. \label{spal_scheme}}

%--- overall comparison TGSR data and simulation ---------------------

\figcaption[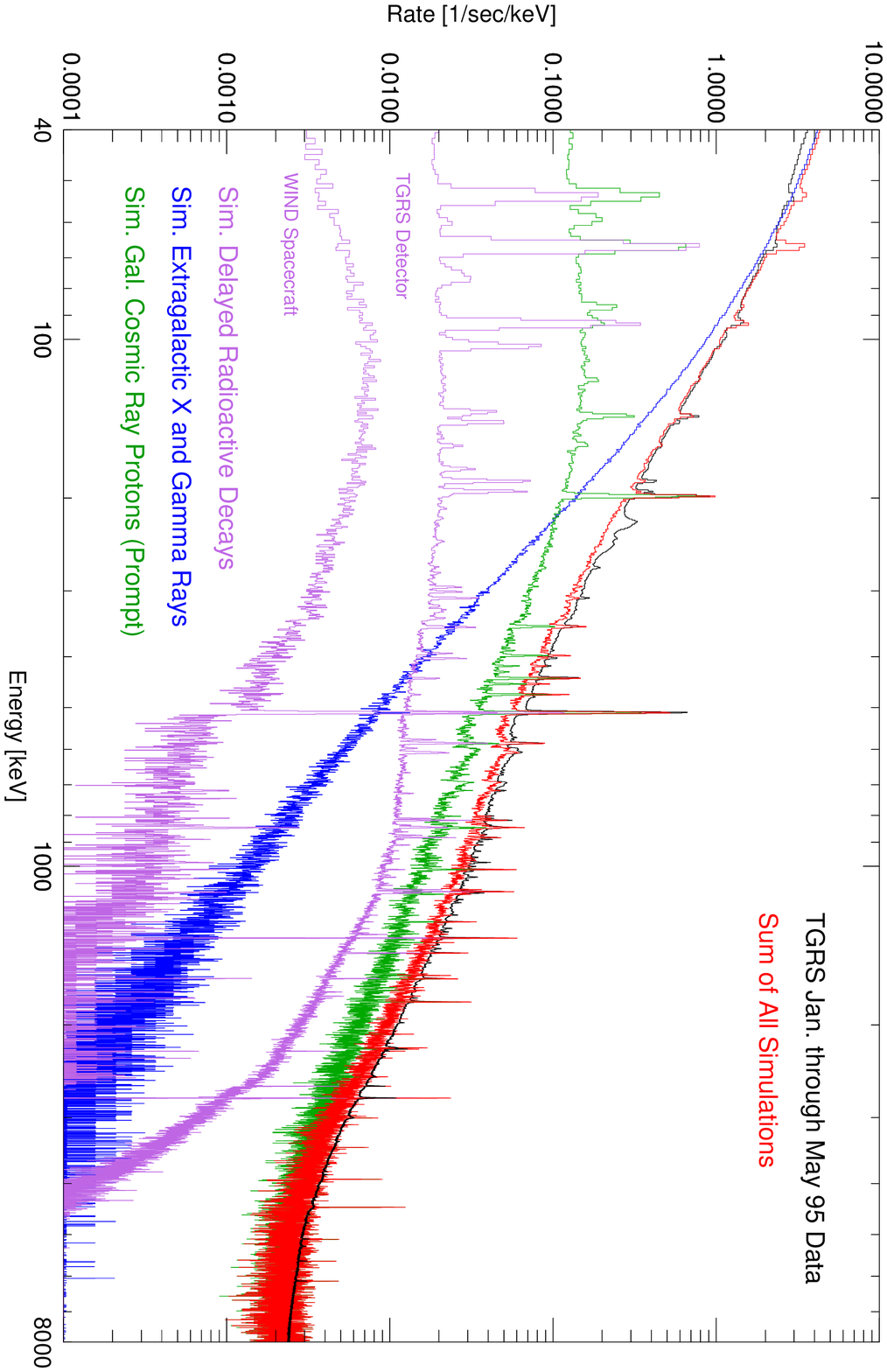]{A comparison of the overall
Jan.--May, 1995 TGRS spectrum with the various components of our MGGPOD
simulation. Details are given in the text. The broad features in the
data in 210--260~keV are electronic artifacts.
\label{compare_tgrs_sim_spc}}

%--- comparison TGSR data and simulation in 40-800 keV ---------------

\figcaption[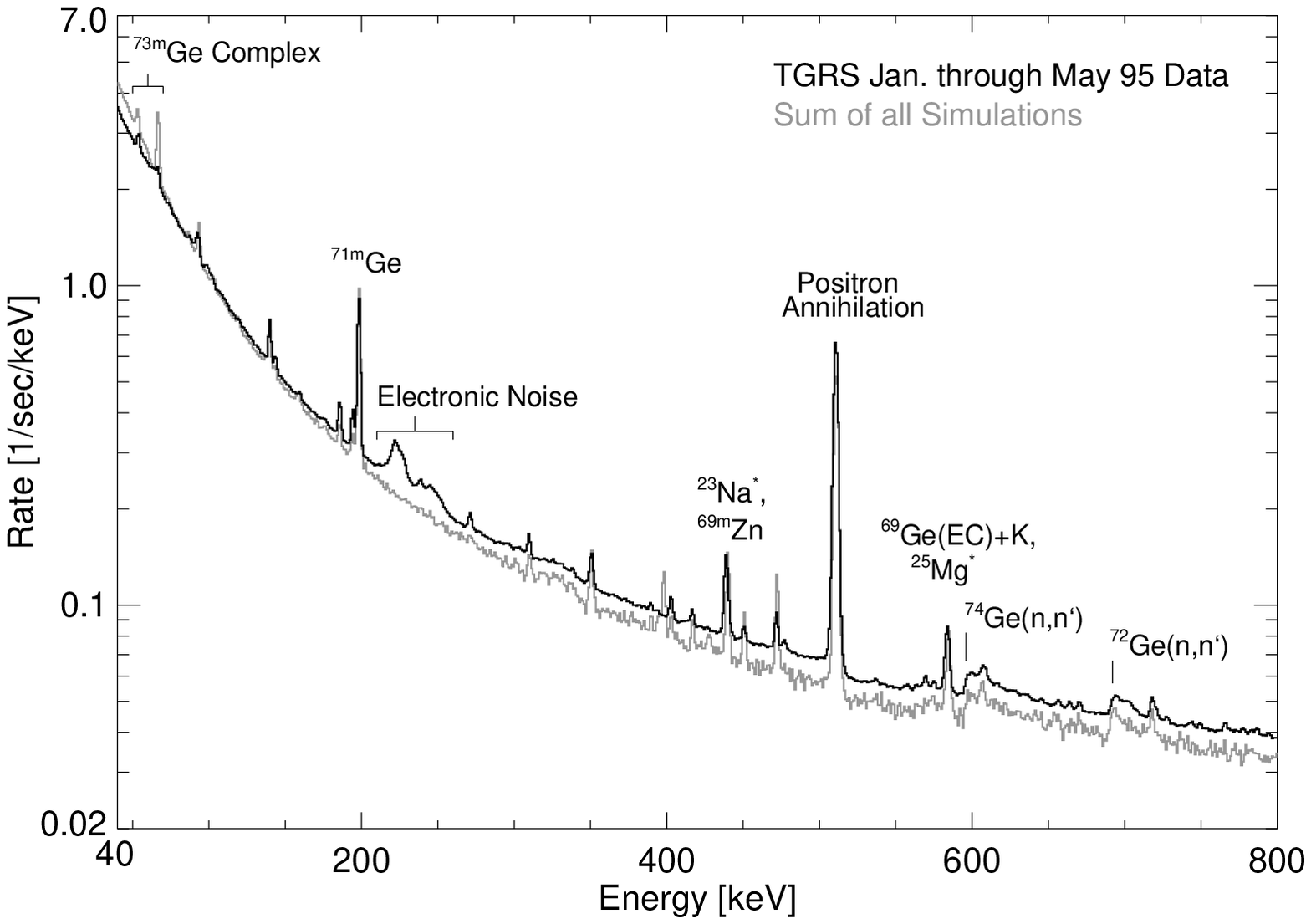]{A comparison of
the Jan.--May, 1995 TGRS spectrum in 40--800~keV with the summed
MGGPOD simulation. Some of the more prominent lines and spectral
features have been labelled, including the inelastic neutron
scattering features corresponding to the 596~keV and 692~keV levels of
$^{74}$Ge and $^{72}$Ge, respectively.
\label{compare_tgrs_sim_spc_40-800}}

%--- comparison TGSR data and simulation in 800-1100 keV --------------

\figcaption[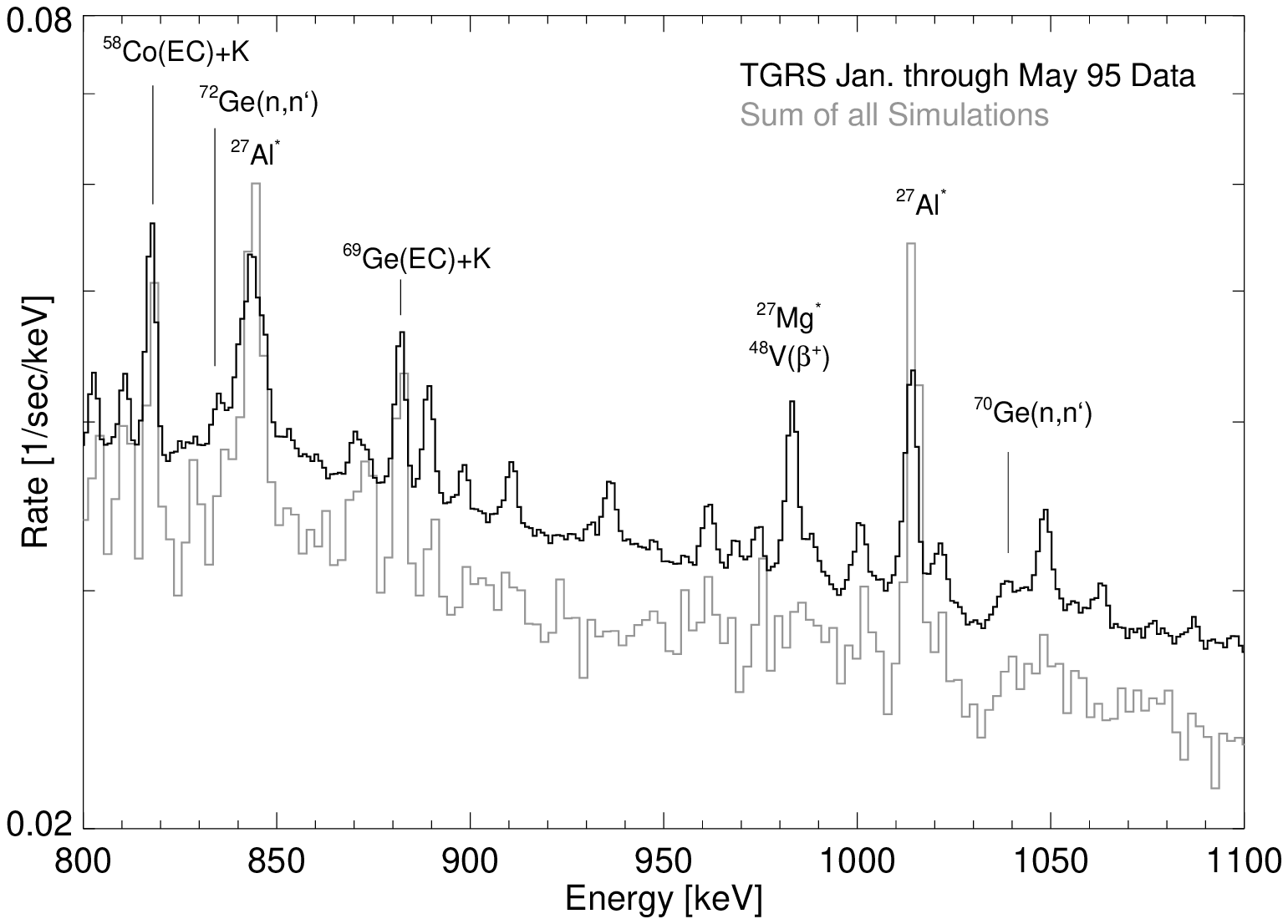]{A comparison of
the Jan.--May, 1995 TGRS spectrum in 800--1100~keV with the summed MGGPOD
simulation. Some of the more prominent lines and spectral
features have been labelled, including the inelastic neutron
scattering features corresponding to the 834~keV and 1039~keV levels of
$^{72}$Ge and $^{70}$Ge, respectively.
\label{compare_tgrs_sim_spc_800-1100}}

%--- comparison TGSR data and simulation in 1100-1900 keV -------------

\figcaption[78.eps]{A comparison of
the Jan.--May, 1995 TGRS spectrum in 1100--1900~keV with the summed MGGPOD
simulation.  Some of the more prominent lines and spectral
features have been labelled, including the inelastic neutron
scattering features corresponding to the 1204~keV and 1215~keV levels of
$^{74}$Ge and $^{70}$Ge, respectively.
\label{compare_tgrs_sim_spc_1100-1900}}

%--- comparison TGSR data and simulation in 1900-3200 keV -------------

\figcaption[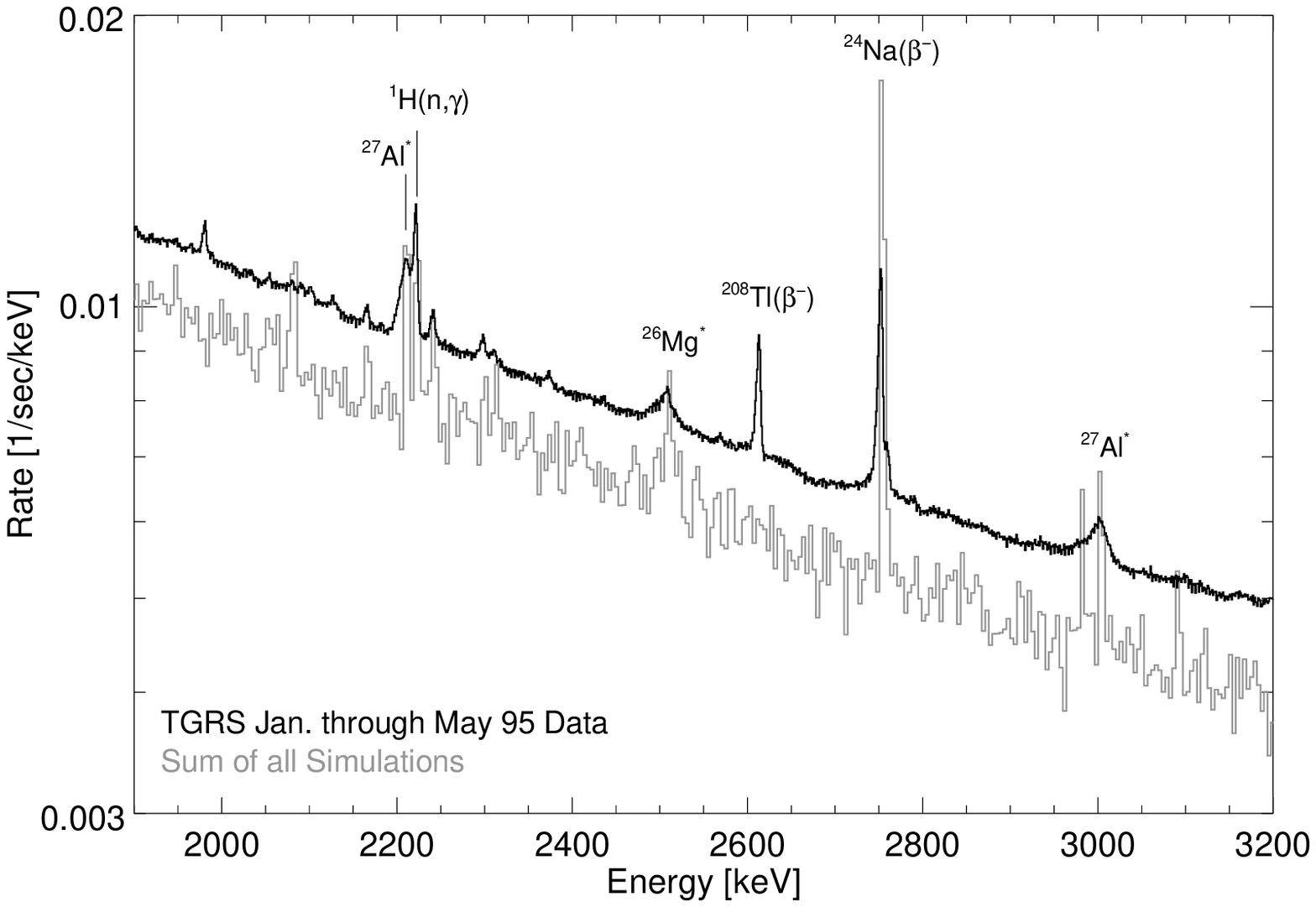]{A comparison of
the Jan.--May, 1995 TGRS spectrum in 1900--3200~keV with the summed MGGPOD
simulation. Some of the more prominent lines and spectral
features have been labelled.
\label{compare_tgrs_sim_spc_1900-3200}}

%--- comparison TGSR data and simulation in 3200-8000 keV -------------

\figcaption[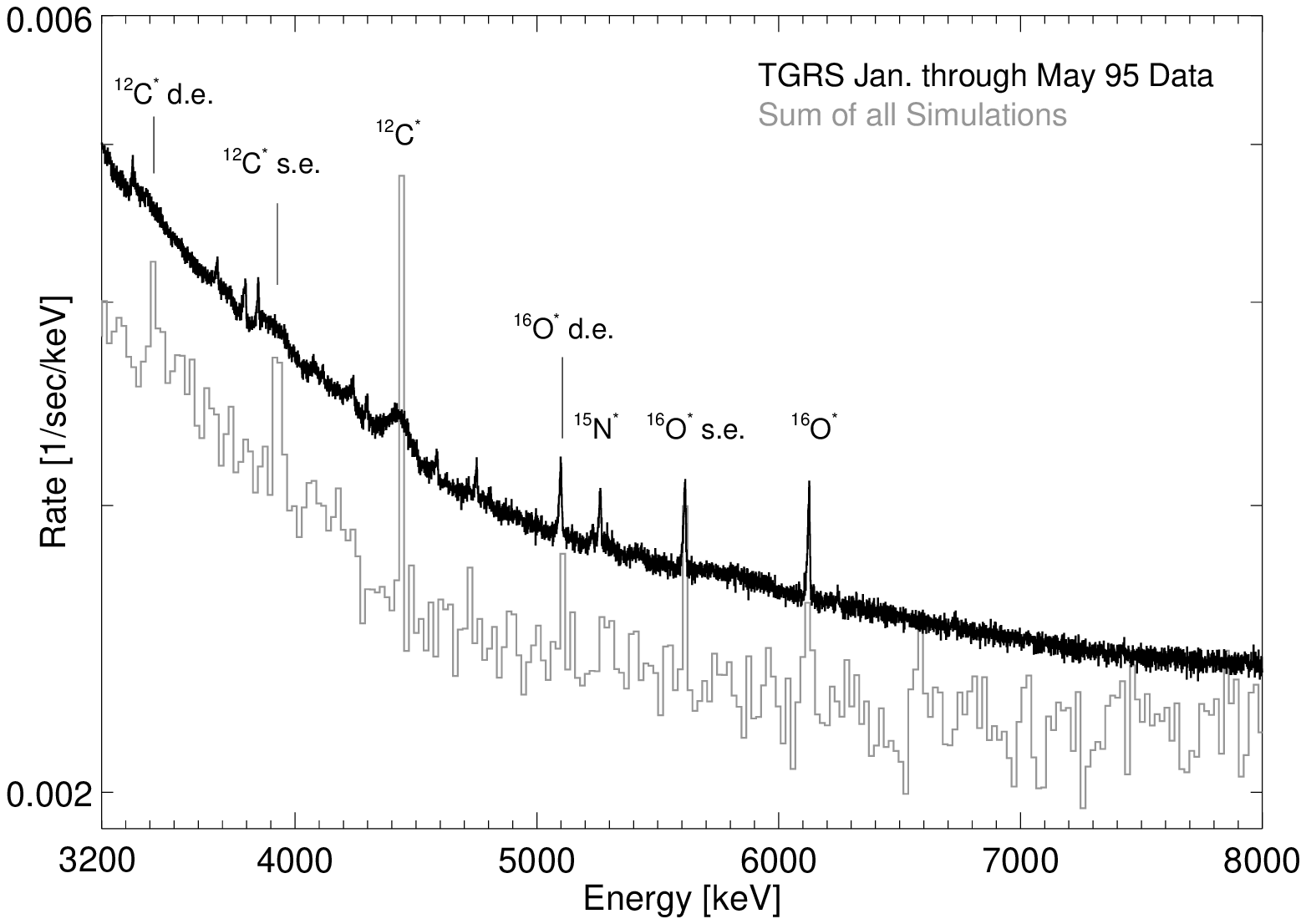]{A comparison of
the Jan.--May, 1995 TGRS spectrum in 3200--8000~keV with the summed
MGGPOD simulation. Some of the more prominent lines and spectral
features have been labelled. The Doppler broadening of e.g.\ the
$^{12}$C$^\ast$ lines is not included in the simulation.
\label{compare_tgrs_sim_spc_3200-8000}}

%--- ratio of simulation and TGRS data -------------------------------

\figcaption[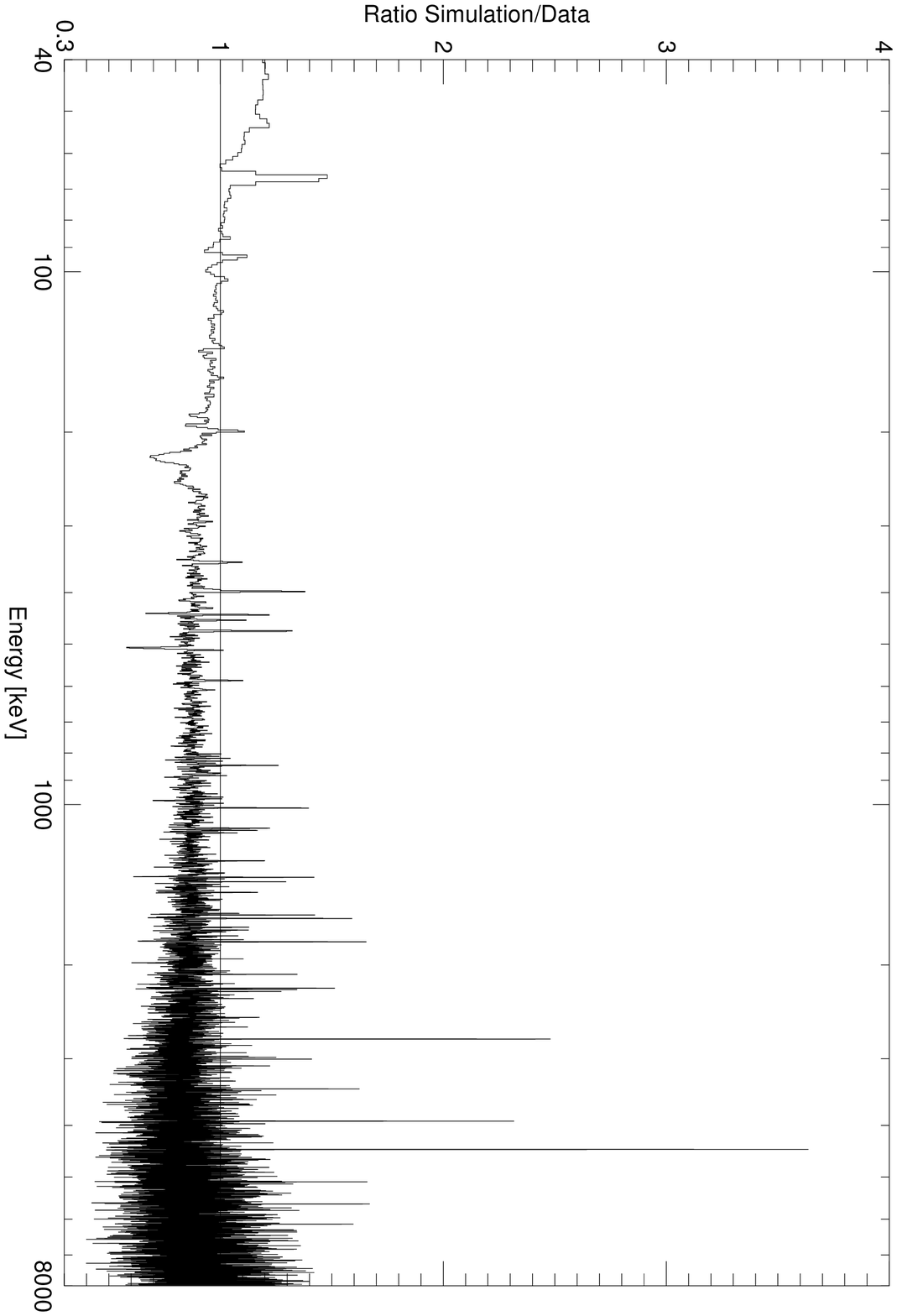]{The ratio of the
MGGPOD simulation and the Jan.--May, 1995 TGRS spectrum.
\label{compare_tgrs_sim_spc_ratio}}

%--- ratio sim/data scatter plot -------------------------------------

\figcaption[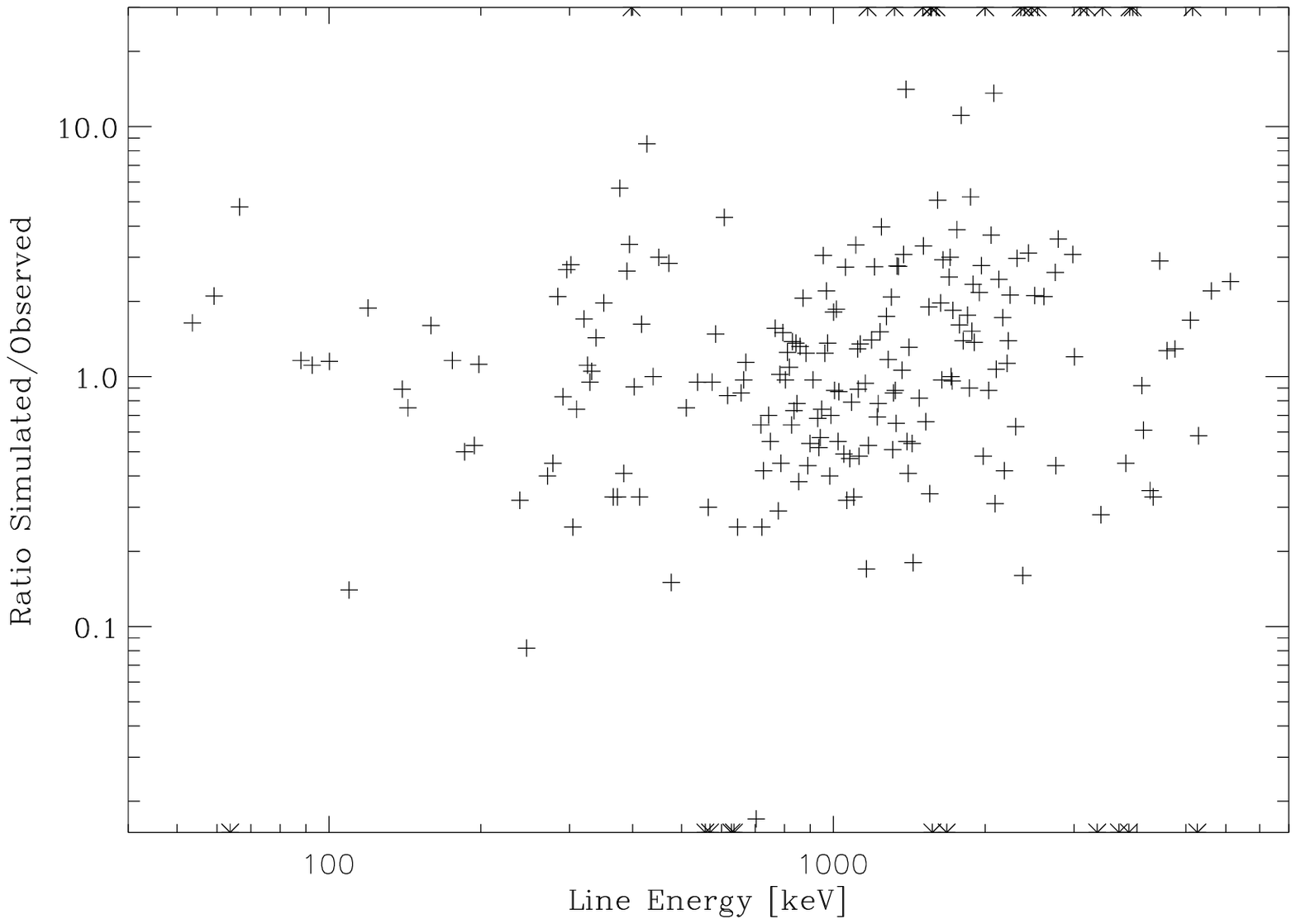]{Ratio between simulated and observed
line strengths for TGRS background spectrum accumulated during
Jan.--May, 1995.  Top arrows -- lines predicted by simulation but not
observed.  Bottom arrows -- observed lines without simulated
equivalents (mostly from $A < 20$ spallation product nuclei not
included in simulation). The lines are well reproduced, typically to
within a factor of 2.5. \label{ratio_simdat_scat_plot}}

%--- ratio sim/data histogram plot -----------------------------------

\figcaption[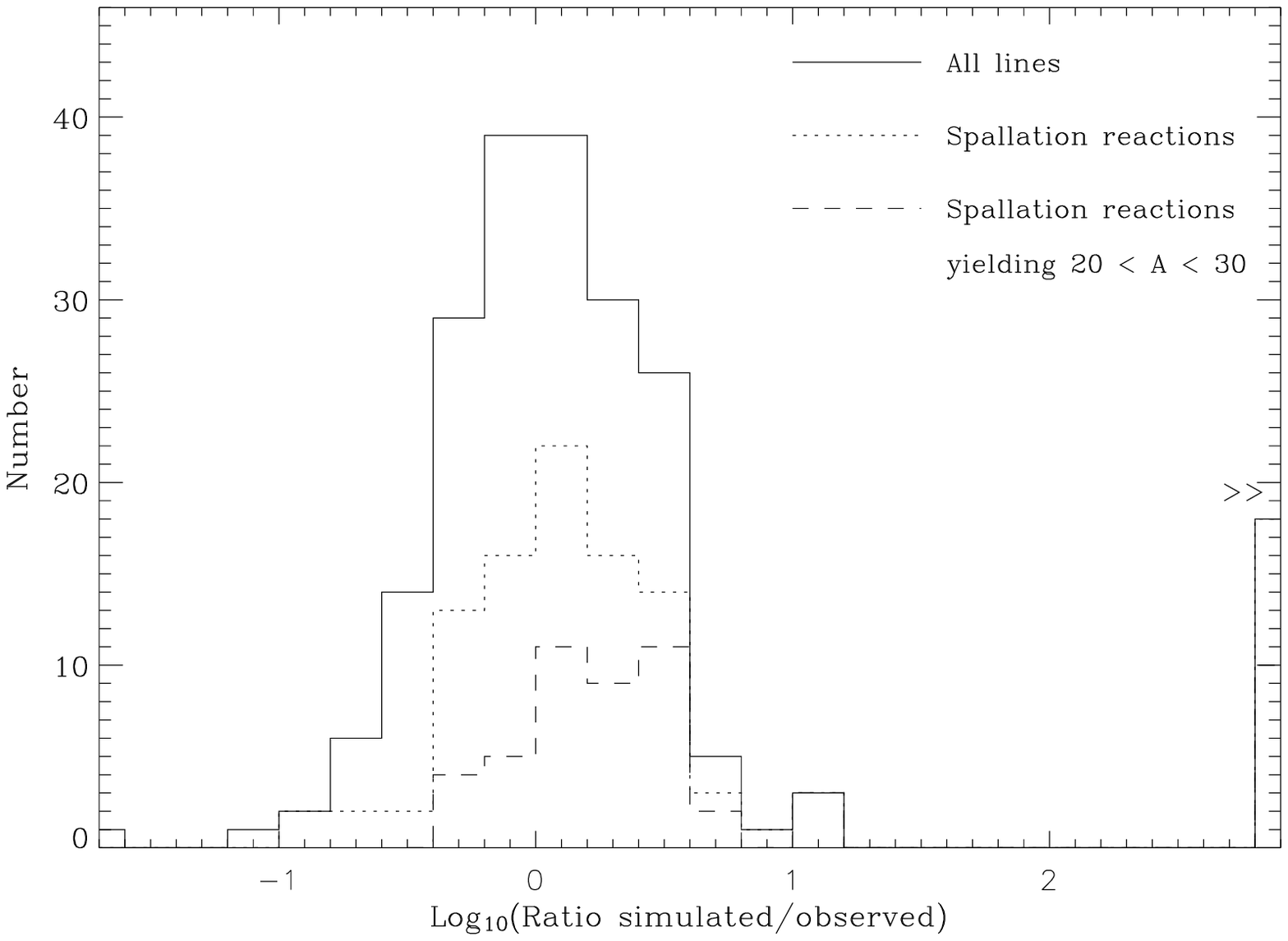]{Distribution of ratios of
simulated to observed line strengths.  Full histogram -- all
lines. Dotted histogram -- de-excitation lines from spallation
reactions.  Dashed histogram -- de-excitation lines from spallation
reactions yielding nuclei with mass $20 \le A \le 30$.  The symbol ``$>>$''
indicates the bin containing lines predicted but not observed
(infinite ratio). \label{ratio_simdat_hist_plot}}

%\figcaption[sgi9259.eps]{This is the first figure and it uses sgi9259.eps as
%its EPS figure file. \label{fig1}}
%
%\figcaption[sgi9279.eps]{This is an example of a long figure caption that
%must be set as a paragraph.  The processor has to buffer the text of the
%caption, so it is good not to be too wordy, but that would make for
%poor communication as well. \label{fig2}}
%
%\figcaption{This figure has no associated EPS file, so the optional
%parameter is omitted. \label{fig3}}

%%%UCP%%%
\newpage
\plotone{f1.eps}
\newpage
\plotone{f2.eps}
\newpage
\plotone{f3.eps}
\newpage
\plotone{f4.eps}
\newpage
\epsscale{.90}
\plotone{f5.eps}
\newpage
\plotone{f6.eps}
\newpage
\plotone{f7.eps}
\newpage
\plotone{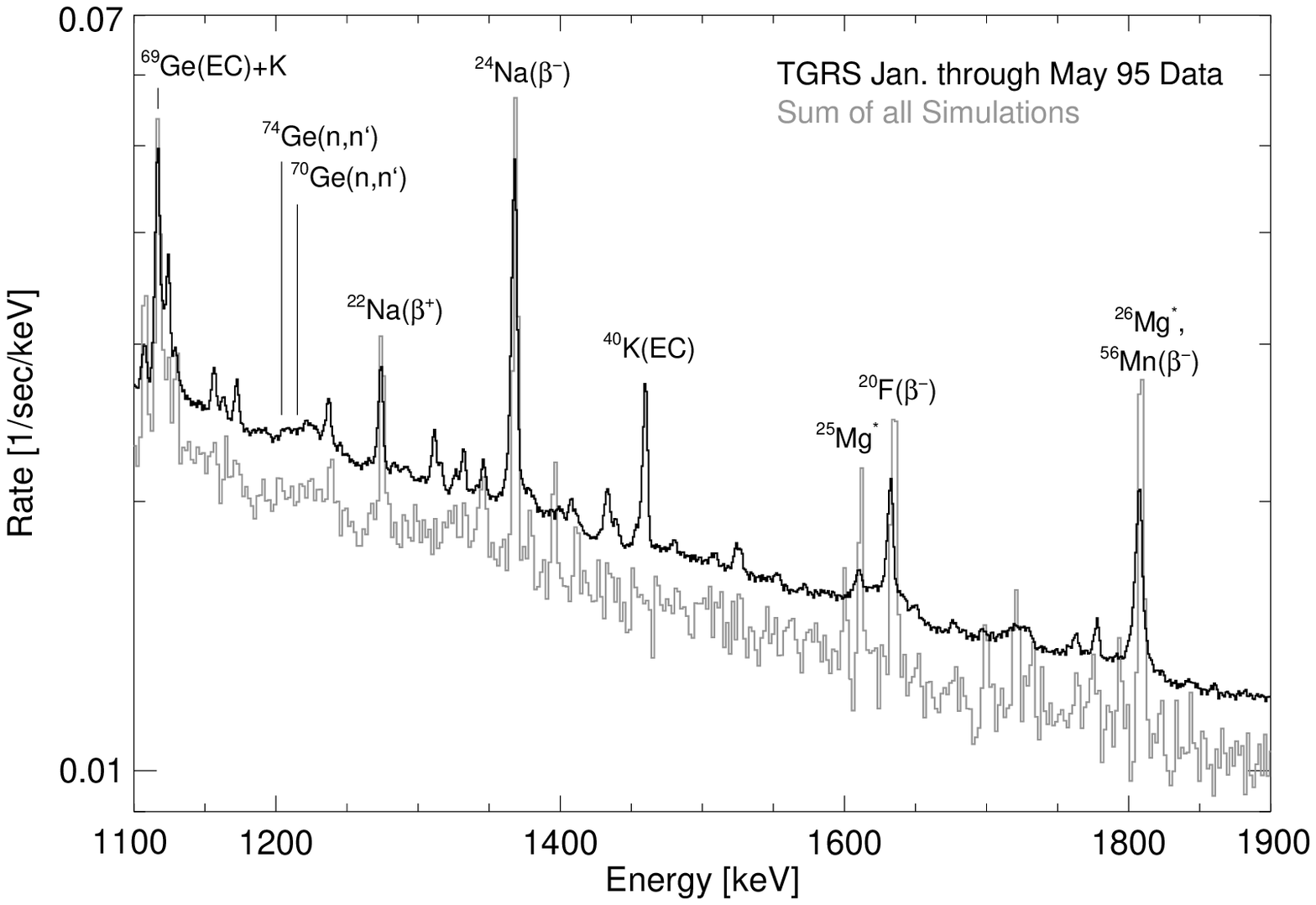}
\newpage
\plotone{f9.eps}
\newpage
\plotone{f10.eps}
\newpage
\plotone{f11.eps}
\newpage
\epsscale{1.0}
\plotone{f12.eps}
\newpage
\plotone{f13.eps}

%------------------------
%         tables
%------------------------

%% Tables should be submitted one per page, so put a \clearpage before

%% Two options are available to the author for producing tables:  the
%% deluxetable environment provided by the AASTeX package or the LaTeX
%% table environment.  Use of deluxetable is preferred.
%%

\clearpage

%--- radiation environment overview ------------------------------

\begin{deluxetable}{lccc}
\tabletypesize{\footnotesize}
\tablecaption{Overview of various components of the space radiation
environment discussed in more detail in the text. For each component,
the main particle species are listed, together with a crude
characterization of the energy spectrum, and an indication of the
orbits for which they are most relevant.
\label{rad_env_table} 
}
\tablewidth{0pt} 
\tablehead{
\colhead{Radiation Environment} & \colhead{Particle}
& \colhead{Energy Spectrum\tablenotemark{a}} & \colhead{Orbits
Affected\tablenotemark{c}} \\ 
\colhead{Component} & \colhead{Species} & \colhead{and
Range\tablenotemark{b}} &  
}

\startdata
%
%Galactic Cosmic Rays & p$^+$, He$^{2+}$, e$^-$, \dots & TBD &
%all orbits \\
%Galactic Cosmic Rays & p & $F \propto E^{-2.7}$, $E \lesssim 1$~TeV &
%All \\
Galactic Cosmic Rays & p, $\alpha$ & $E^{-2.7}$, $1~\mbox{GeV} - 1~\mbox{TeV}$\tablenotemark{d} &  All \\ 
%                     & e & $F \propto E^{-2.6}$, $E \lesssim 100$~GeV &
%All \\
                     & e & $E^{-2.6}$, $1-100$~GeV\tablenotemark{d}  & All \\ 
%Solar and Anomalous Cosmic Rays & p$^+$, \dots & TBD & outside
%magnetosphere \\
Solar Energetic Particles & p & $E^{-3.3}$, $E < 200$~MeV &
HEO, Balloon (LC) \\
Anomalous Cosmic Rays & He$^+$, O$^+$, Ne$^+$, \dots  & 10--100~MeV &
HEO, Balloon (LC) \\
%Geomagn.\ Trapped Rad.\ & p$^+$, e$^-$, \dots & TBD & low-Earth
%orbits \\
Geomagn.\ Trapped Rad.\ & p & $E^{-0.25}$, $E <
400$~MeV & LEO \\
                        & e$^-$ & $E^{-4}$, $E <
10$~MeV & LEO \\
%Earth Albedo Radiation & $\gamma$, n, \dots & TBD & Balloon, LEO \\
Earth Albedo Radiation & $\gamma$ & $E^{-1.7}$, $10~\mbox{keV} -
1~\mbox{GeV}$ & Balloon, LEO \\ 
  & n & $E^{-1.3}$, thermal -- 10~GeV &
Balloon, LEO \\ 
Diffuse Cosmic Photons & $\gamma$ & $E^{-2.3}$, 50~keV -- 1~GeV\tablenotemark{e}
& All \\ 
%Intern.\ Prod.\ Background & $\gamma$, n, p, e$^\pm$,
%$\mu^\pm$, $\pi^{0,\pm}$ ,\dots & TBD & All\\ 
%  & radioactive nuclei & &
Intern.\ Prod.\ Background & $\gamma$ & cont.\ \&
lines (10~keV--10~MeV)\tablenotemark{f} & All \\ 
                           & n, p, e$^\pm$ &
complex\tablenotemark{f} & All \\ 
\enddata
\tablenotetext{a}{ Energy spectra are crudely approximated by power laws in
order to illustrate the "hardnesses" involved. The spectra are
differential intensities in units of particles per energy (and per
nucleon in case of hadrons) per unit solid angle, time, and area. Energies
are kinetic energy, except in the case of photons.}
\tablenotetext{b}{ The approximate range of energies covered by these
spectra, as far as they are relevant for the production of
instrumental background. 
%Energy ranges are sometimes limit by low and/or high energy
%cut-offs.
Note that the upper bounds do not necessarily indicate where the
spectrum comes to an end, but where it becomes of negligible
importance. The lower bounds result from complex physical processes
described in the following notes.}  
\tablenotetext{c}{ Key to (idealized) orbits: Balloon --- balloon
altitude, near top of atmosphere; Balloon (LC) --- as Balloon, but
specifically at low geomagnetic cut-off (i.e.\ at high geographic
latitudes); LEO --- low-Earth orbit (will intersect radiation belts);
HEO --- high-Earth orbit (outside Earth's magnetosphere).}
\tablenotetext{d}{ The lower bounds and the low energy spectral shape
are strongly dependent on the phase of the solar cycle (solar
modulation), and for LEO even more so on orbit altitude and
inclination (geomagnetic cut-off).}
\tablenotetext{e}{ At lower X-ray energies, the spectrum steepens to
$E^{-1.4}$.}
\tablenotetext{f}{ The $\gamma$ and particle spectra are very complex
-- the calculations thereof being the theme of this paper -- and
heavily dependent on the design of the instrument and the radiation
environment in which it is operated.}

\end{deluxetable}

%--- MGGPOD simulation steps -------------------------------------

\begin{deluxetable}{ccc}
\tabletypesize{\footnotesize}
\tablecaption{The sequence of simulation steps for Class~I and
Class~II radiation fields. Details are explained in the text. A flow
chart of MGGPOD is given in Fig.~\ref{mggpod_flow_chart}.
\label{mggpod_sim_steps} 
}
\tablewidth{0pt} 
\tablehead{
\colhead{Step 1 (Class~I and II)} & \colhead{Step 2 (Class II only)} &
\colhead{Step 3 (Class II only)}
} 

\startdata
Mass Model \&  & Isotope Production Rates & Mass Model \&  \\
Radiation Environment & & Activity per Isotope and Volume \\
{\Large $\downarrow$} & {\Large $\downarrow$} & {\Large $\downarrow$} \\
MGEANT, GCALOR, & ORIHET & MGEANT, GCALOR,  \\
PROMPT &  & DECAY \\
{\Large $\downarrow$} & {\Large $\downarrow$} & {\Large $\downarrow$} \\
Prompt Energy Deposits &  &  \\ 
(Class~I and II), & Activity per Isotope & Delayed Energy Deposits \\
\& Isotope Prod.\ Rates & and Volume & \\
(Class~II) & & \\
\enddata

\end{deluxetable}

%--- neutron capture table ---------------------------------------

\begin{deluxetable}{ll}
\tabletypesize{\small}
\tablecaption{Isotopes for which prompt de-excitation photons are
generated after neutron capture and/or inelastic neutron
scattering. \label{n_isot_table}
}
\tablewidth{0pt} 
\tablehead{
\colhead{Process} & \colhead{Product Isotopes} 
} 

\startdata
Neutron Capture $(n,\g)$ & $^{10}$Be, $^{28}$Al, $^{49}$Ti, $^{53}$Cr,
$^{55,57}$Fe, $^{64,66}$Cu, $^{71,73,74,75,77}$Ge \\ 
Inelastic Neutron Scattering $(n,n^\prime\g)$ & $^{27}$Al,
$^{54,56,57}$Fe, $^{63,65}$Cu, $^{70,72,73,74,76}$Ge \\ 
\enddata

\end{deluxetable}

%--- ``Ramaty nuclei'' ----------------------------------------------

\begin{deluxetable}{ll}
\tablecaption{Nuclei produced by spallation for which de-excitation line
probabilities were obtained from \cite{Ramaty79}.
\label{ramaty_isot_table}
}
\tablewidth{0pt} 
\tablehead{
\colhead{Process} & \colhead{Product Isotopes} 
}

\startdata
Spallation ($A < 20$) & $^{10}$B, $^{11,12}$C, $^{14}$N, $^{15,16}$O, 
$^{19}$F \\
\enddata

\end{deluxetable}

%--- class 1 lines table ---------------------------------------------

\begin{deluxetable}{lclcl}
\tabletypesize{\scriptsize}
\tablecaption{Identified lines; blends of identified lines.
\label{class1_ids}
} 

\tablewidth{0pt}
\tablehead{
\colhead{TGRS line} & \colhead{Count rate} & \colhead{Transition ID} &
\colhead{Lab} & \colhead{Comment\tablenotemark{a}} \\ 
\colhead{energy\tablenotemark{b}} & \colhead{s$^{-1}$} &
\colhead{(levels)\tablenotemark{b,c}} &
\colhead{energy\tablenotemark{b}} & \colhead{} 
}

\startdata

53.6    &    0.71  & $^{73m}$Ge(67--13) & 53.4 & $a$,$s$ 
~$\sim 15$\% buildup on $^{73}$As $\tau_{1/2}$. \\
 &            &   $^{65m}$Zn(54--g.s.) & 53.9 & $a$ ~~~From
$^{65}$Ga($\beta^+$). \\ 
  &                      &   $^{58}$Co(53--g.s.) & 53.0 & $s$  \\
    &                    &   $^{46}$Sc(52--g.s.) & 52.0 & $s$   \\
59.2 &     0.100   &          $^{60m}$Co(59--g.s.) & 58.6 & $s$  \\ 
      &      &             ? $^{74m}$Ga(60--g.s.)  & 59.7 & $s$  \\
66.5   &    0.65  &  $^{73m}$Ge(67--13--g.s.) & 66.7 & $a$,$s$ ~$\sim 5$\% 
buildup on $^{73}$As $\tau_{1/2}$. \\
    &        &            &    & ~~~~~Two-step transition. \\
88.0    &  0.25          &  $^{69m}$Ge(87--g.s.) & 86.8  & $a$,$s$ ~From 
$^{69}$As($\beta^+$)  \\
      &         &     $^{24}$Na(563--472) & 91.0 & $s$  \\
92.6    &    0.88      &    $^{67m}$Zn(93--g.s.) & 93.3 & $a$,$s$ ~From 
$^{67}$Ga(EC).  \\
   &        &    $^{67m}$Zn(93--g.s.) + L  & 94.5 & $a$ ~~~From 
$^{67}$Ga(EC).    \\
119.5  &     0.051    &       $^{72m}$Ga(120--16--g.s.)  & 119.5 & $a$,$s$ 
  ~~Two-step transition. \\
139.7   &    0.47     &    $^{75m}$Ge(140--g.s.) & 139.7 & $a$,$s$ ~From 
$^{75}$Ga($\beta^-$). \\
         &          &  ? $^{57}$Fe(136--14--g.s.) + L  & 137.3 &  $a$ ~~~From
$^{57}$Co(EC). \\
143.4   &    0.13  &   $^{57}$Fe(136--14--g.s.) + K & 143.6 & $a$ ~~~60\%
buildup on $^{57}$Co $\tau_{1/2}$.  \\
     &          &         $^{46m}$Sc(143--g.s.) & 142.5 & $s$  \\
159.3     &    0.055    &     $^{47}$Ti(159--g.s.) & 159.4 & $a$,$s$ ~From
$^{47}$Sc($\beta^-$).  \\
 &                     &     $^{68}$As(158--g.s.) & 158.1 & $s$   \\
175.7    &     0.031      &   $^{71}$Ge(175--g.s.) + L & 176.4 & $a$,$s$ ~From
$^{71}$As(EC). \\
         &               &   $^{71}$Ge(175--g.s.) &  174.9 & $n$  \\
185.7  &      0.26     &   $^{67}$Zn(185--g.s.) + L & 185.8 & $a$ ~~~From 
$^{67}$Ga(EC). \\      
      &  &  $^{71}$Ge(175--g.s.) + K & 186.1 & $a$ ~~~From $^{71}$As(EC).  \\
194.2  &      0.24     &  $^{67}$Zn(185--g.s.) + K & 194.2  & $a$ ~~~From 
$^{67}$Ga(EC).  \\  
198.2  &   1.47 &    $^{71m}$Ge(198--175--g.s.)  & 198.4 & $a$,$s$ ~From 
$^{71}$As($\beta^+$). Two-step transition. \\
   &     &    $^{19}$F(197--g.s.)  & 197.1&  $s$,$a$ ~$\sim 2$\% buildup 
follows solar modulation.  \\
239.1   &   0.165  &  $^{212}$Bi(239--g.s.) & 238.6 & $r$
~~~$^{232}$Th series. \\ 
271.2  &     0.053    &     $^{44m}$Sc(271--g.s.) & 271.1 & $s$   \\
     &      &           $^{228}$Th(328--58) & 270.2 & $r$
~~~$^{232}$Th series. \\ 
278.0   & 0.026   &  $^{208}$Pb(3475--3198)  & 277.4 & $r$
~~~$^{232}$Th series. \\ 
       &        &       $^{228}$Th(1153--874)   & 279.0 & $r$ ~~~$^{232}$Th 
series.  \\
284.2    &     0.0023   &  $^{61}$Ni(283--g.s.) + L  & 284.0 &  $a$ ~~~From 
$^{61}$Cu(EC).  \\
291.2    &     0.0046   &   $^{61}$Ni(283--g.s.) + K & 291.3 & $a$ ~~~From 
$^{61}$Cu(EC).  \\
296.1  & 0.0153    &     $^{214}$Bi(295--g.s) & 295.2 & $r$
~~~$^{238}$U series \\ 
301.8   &    0.0082    &     $^{67}$Zn(394--93) + L  & 301.4 & $a$ ~~~From 
$^{67}$Ga(EC).  \\
304.5   & 0.015  &   $^{75m}$As(304--g.s.)   & 303.9 & $s$   \\
309.6  &     0.062    &    $^{67}$Zn(394--93) + K  & 309.9 & $a$ ~~~From 
$^{67}$Ga(EC).  \\    
320.4   &    0.0056    &      $^{51}$V(320--g.s.)  & 320.1 & $a$,$s$ ~From 
$^{51}$Ti($\beta^-$), $^{51}$Cr(EC). \\
     &       &       $^{228}$Th(1154--832)  & 321.6 & $r$
~~~$^{232}$Th series \\ 
325.6    &     0.0103    &   $^{51}$V(320--g.s.) + K & 325.6 & $a$ ~~~From
$^{51}$Cr(EC). \\
329.2      & 0.0056    &       $^{228}$Th(328--g.s.)  & 328.0 & $r$
~~~$^{232}$Th  
series.  \\
    &   &  ? $^{69}$Ga(319--g.s.) + K & 329.0 & $a$ ~~~From $^{69}$Ge(EC).  \\
331.9   &  0.0076   &        $^{21}$Na(332--gs) & 331.9 & $s$   \\
338.5 &    0.013 &       $^{59}$Ni(339--g.s) & 339.4 & $s$   \\
          &           &        $^{228}$Th(396--58) & 338.3 & $r$
~~~$^{232}$Th series.  \\
350.7 &      0.088   &      $^{21}$Ne(350--g.s.)  & 350.7 & $s$,$a$ ~From 
$^{21}$Na($\beta^+$).  \\
    &             &  $^{214}$Bi(352--g.s.) & 351.9 & $r$ ~~~$^{238}$U
series.  \\ 
373.2     &    0.0066    &    $^{43}$Ca(373--g.s.)  & 372.8 &  $a$ ~~~From 
$^{43}$K($\beta^-$), $^{43}$Sc(EC).\\
377.1   &    0.0024   &       $^{52m}$Mn(378--g.s.)  & 377.7 &  $s$   \\
389.8   &    0.0110  &      $^{25}$Mg(975--585) & 389.7 & $s$,$a$ ~From 
$^{25}$Na($\beta^-$). \\
       &      &    $^{214}$Po(1764--1378)  & 387.0 &  $r$ ~~~$^{238}$U
series. \\ 
      &         &    $^{214}$Po(2119--1730)  & 389.1 &  $r$
~~~$^{238}$U series. \\ 
394.2   &    0.0077    & $^{67}$Zn(394--g.s.) + L & 394.7 & $a$ ~~~From 
$^{67}$Ga(EC).  \\
402.9  &    0.044  &   $^{67}$Zn(394--g.s.) + K & 403.2 &  $a$ ~~~From 
$^{67}$Ga(EC).  \\ 
416.8    &  0.026    &     $^{26}$Al(417--g.s.) & 416.9 & $s$  \\
426.9    &   0.0034   &        $^{73m}$As(428--g.s.)  & 428.3 & $s$  \\
439.2 &    0.21 &  $^{23}$Na(440--g.s.)  & 440.0 &  $s$,$a$ ~From 
$^{23}$Ne($\beta^-$), $^{23}$Mg($\beta^+$).  \\
   &           &      $^{69m}$Zn(439--g.s.) & 438.6 & $s$  \\
450.8  &     0.021   &     $^{23}$Mg(451--g.s.)  & 450.7 &  $s$  \\
     &           &       $^{25}$Al(452--g.s.) & 451.5 & $s$ \\
472.2   &    0.056   &     $^{24m}$Na(472--g.s.)  & 472.2 & $s$  \\
477.1    &   0.022     &   $^{7}$Li(478--g.s.)    & 477.6 & $s$    \\
511.0     &    2.26   &  $e^{+}$ annihilation & 511.0  & $s$,$a$ 
~$\sim 2$\% buildup follows solar modulation, \\
  &   &   &   & \phantom{$s$,$a$} kinematically broadened. \\
538.0    &  0.0042         & $^{59}$Fe(1750--1211) & 537.4  & $s$     \\
564.8    &     0.0089  &   $^{228}$Th(1531--969) & 562.5 &  $r$ ~~~$^{232}$Th
series \\
 & &       $^{54}$Cr(3786--3222) & 563.7 & $a$ ~~~From $^{54}$V($\beta^-$). \\
575.1     &  0.021    &    $^{69}$Ga(574--g.s.) + L & 575.4 &  $a$ ~~~From 
$^{69}$Ge(EC). \\
584.0    &   0.125   &     $^{69}$Ga(574--g.s.) + K & 584.5 & $a$ ~~~From 
$^{69}$Ge(EC).  \\
       &      &            $^{25}$Mg(585--g.s.)  & 585.0 &  $s$    \\
       &      &            $^{22}$Na(583--g.s.)  & 583.0 & $s$,$a$ ~From 
$^{22}$Mg($\beta^+$).  \\
     &      &              $^{208}$Pb(3198--2615)  & 583.2 & $r$ ~~~$^{232}$Th
series.  \\
608.0  & 0.030 &        $^{74}$Ge(596--g.s.) + K & 606.9 & $a$ ~~~From 
$^{74}$As(EC).  \\
     &        &          $^{214}$Po(609--g.s.)   & 609.3 & $r$ 
~~~$^{238}$U series.  \\
656.7   &    0.0067  &     $^{20}$F(656--g.s.) & 656.0 & $s$   \\
   &    &              $^{61}$Ni(656--g.s.) + L & 657.0 & $a$ ~~~From
$^{61}$Cu(EC). \\
  &    &   $^{61}$Ni(656--g.s.) & 656.0  &  $s$ \\
664.2 &      0.0070 &       $^{61}$Ni(656--g.s.) + K & 664.3 & $a$ ~~~From
$^{61}$Cu(EC). \\
          &    &      $^{214}$Po(1275--609) & 665.5 & $r$ ~~~$^{238}$U
series  \\ 
670.4  &      0.0103    &     $^{63}$Cu(670--g.s.)  & 669.6 &   $s$   \\
  &  &                 $^{38m}$Cl(671--g.s.) & 671.4 & $s$   \\  
718.4 &      0.025     &   $^{10}$B(718--g.s.)  & 718.3 & $s$   \\
721.4   &    0.0111   &          $^{46}$Sc(774--52)  & 721.9 & $s$  \\
727.2  &     0.0101 &       $^{212}$Po(727--g.s.)   & 727.3 & $r$
~~~$^{232}$Th series. \\
    &         &          $^{228}$Th(1123--396)    & 726.9 &  $r$ ~~~$^{232}$Th
series. \\
744.0  &     0.0069  &      $^{52}$Cr(3114--2370) & 744.4 & $a$ ~~~From 
$^{52}$Mn(EC).  \\
       &            &        $^{234}$U(786--43) & 742.8 & $r$ ~~~$^{238}$U 
series. \\
749.9  &     0.0060 &       $^{56}$Co(1720--970)  & 750.0 & $a$ ~~~From 
$^{56}$Ni(EC).  \\
767.0 &     0.0115  &      $^{234}$U(810--43) & 766.4 & $r$ ~~~$^{238}$U 
series.  \\
 & &                      $^{214}$Po(1378--609)  & 768.4 & $r$ ~~~$^{238}$U 
series. \\
783.3  &  0.0041  &   $^{24}$Na(1345--563)  & 781.4 & $s$  \\
       &          &   $^{50}$Cr(783--g.s.)  & 783.3 & $s$  \\
794.5     &    0.0036     &   $^{228}$Th(1123--328) & 794.9 & $r$ 
~~~$^{232}$Th series. \\
       &              &       $^{27}$Al(3004--2211) & 793.0 & $s$ \\
803.4   &    0.0153  &      $^{206}$Pb(803--g.s.) & 803.1 & $s$   \\
811.2  &    0.0176  & $^{58}$Fe(811--g.s.) + L & 811.6 &  $a$ ~~~50\%
buildup on $^{58}$Co $\tau_{1/2}$. \\
    &      &          $^{58}$Fe(811--g.s.) & 810.8 & $a$,$s$ ~From 
$^{58}$Co(EC). \\
818.2 &      0.055 &  $^{58}$Fe(811--g.s.) + K  & 817.9 & $a$ ~~~66\%
buildup on $^{58}$Co $\tau_{1/2}$. \\
829.9    &   0.011     &       $^{26}$Al(1058--228)  & 829.4 & $s$   \\
835.3  &  0.014 &   $^{54}$Cr(835--g.s.) + L      & 835.5 & $a$ ~~~From
$^{54}$Mn(EC).   \\
  &   &       $^{54}$Cr(835--g.s.)  & 834.8 & $a$ ~~~From
$^{54}$V($\beta^-$). \\ 
  &  &              $^{228}$Th(1023--187) & 835.7 &  $r$ ~~~$^{232}$Th
series.  \\ 
842.7 &    0.080  &  $^{27}$Al(844--g.s.) & 843.7 & $s$,$a$ ~From 
$^{27}$Mg($\beta^-$).  \\
    &      &          $^{54}$Cr(835--g.s.) + K & 840.8 & $a$ ~~~From 
$^{54}$Mn(EC).   \\
847.0  & 0.036   &   $^{56}$Fe(847--g.s.) & 846.8 & $s$,$a$ ~23\%
buildup on $^{56}$Co $\tau_{1/2}$. \\
854.0    &   0.0093  &      $^{56}$Fe(847--g.s.) + K & 853.9 & $a$ 
~$\sim 25$\% buildup on $^{56}$Co $\tau_{1/2}$. \\
859.3      &   0.0068   &      $^{208}$Pb(3475--2615) & 860.6 & $a$
~~~$^{232}$Th series. \\
871.1     & 0.017        &        $^{24}$Na(1341--472) & 869.2  &  $s$  \\
       &          &   $^{24}$Na(1346--472)       & 874.4 &  $s$. \\
     &   &   $^{69}$Ga(872--g.s.) + L & 873.3 & $a$ ~~~From 
$^{69}$Ge(EC).  \\
882.7   &    0.038  &    $^{69}$Ga(872--g.s.) + K & 882.3 & $a$ ~~~From 
$^{69}$Ge(EC). \\
         &            &     ? $^{21}$Na(3680--2798) & 881 & $s$   \\
890.1   &  0.027      &    $^{22}$Na(891--g.s.) & 890.9 & $s$ \\ 
   &     & $^{46}$Ti(889--g.s.) & 889.3 & $a$  ~~10\%
buildup on $^{46}$Sc $\tau_{1/2}$. \\
899.0    &  0.0092   & $^{204}$Pb(899--g.s.) & 899.2 &  $s$  \\
911.3  &     0.0138   &    $^{228}$Th(969--58) & 911.2 &  $r$ ~~~$^{232}$Th 
series. \\
          & &        $^{61}$Ni(909--g.s.)  & 908.6 &  $a$ ~~~From 
$^{61}$Co($\beta^-$)  \\
931.1  &  0.0041   & $^{55}$Fe(931--g.s.)  & 931.3 & $s$  \\
936.5    &   0.0186  &       $^{214}$Po(1543--609) & 934.1 &  $r$
~~~$^{232}$Th series. \\
  &  &     $^{52}$Cr(2370--1434) & 935.5 &  $a$ ~~~From $^{52}$Mn(EC).  \\
   &   &                     ? $^{18}$F(937--g.s.) & 937.2 & $s$  \\
942.3    &   0.0046   &      $^{52}$Cr(2370--1434) + K  & 941.5 &  $a$ ~~~From
$^{52}$Mn(EC).   \\
955.1  &  0.0020      &     $^{27}$Mg(1940--985) & 955.3 &  $s$   \\
962.0  &  0.017  &      $^{63}$Cu(962--g.s.)  & 962.1 &  $s$   \\
            &    &   $^{228}$Th(1023--58)  & 964.8 & $r$ ~~~$^{232}$Th
series.  \\ 
968.8     &   0.0075  &      $^{228}$Th(969--g.s.) & 969.0 &$r$ 
~~~$^{232}$Th series.    \\
974.5     &   0.0121 &       $^{25}$Mg(975--g.s.) & 974.4 & $s$,$a$ ~From
$^{25}$Na($\beta^-$).  \\
983.4   &    0.043  &     $^{27}$Mg(985--g.s.)  & 984.6 & $s$   \\  
      &         &            $^{48}$Ti(984--g.s.) & 983.5 & $s$,$a$  ~~From
$^{48}$V($\beta^+$).  \\
988.9   &      0.0157  &   $^{48}$Ti(984--g.s.) + K & 988.5 & $a$ ~~~From
$^{48}$V(EC).  \\
       &  &                   $^{25}$Mg(1965--975) & 989.9 & $s$  \\
1001.2  &     0.0182 &        $^{234}$U(1045--43) & 1001.0 & $r$ ~~~$^{238}$U
series. \\
          &            &   ? $^{70}$Ga(1003--g.s.) & 1002.6 &  $s$   \\
1006.0  &  0.0033     &   $^{26}$Al(3074--2070)  &  1004.1 & $s$  \\
   &           &           $^{53}$Cr(1006--g.s.)  &  1006.1  & $s$ \\
1014.1  &    0.059  &      $^{27}$Al(1014--g.s.) & 1014.4 &  $s$,$a$ ~From
$^{27}$Mg($\beta^-$) \\
    &           &           $^{26}$Al(2069--1058)  & 1011.7 & $s$   \\
1021.6  &    0.021  &       $^{10}$B(1740--718) & 1021.7  & $s$,$a$ ~From
$^{10}$C($\beta^+$). \\
1025.3  & 0.0030  &  $^{200}$Pb(1027--g.s.)  & 1026.5 &  $s$,$a$ ~~~From 
$^{200}$Bi(EC). \\
1049.3  &    0.020  &      $^{66}$Zn(1039--g.s.) + K & 1048.9 & $a$ ~~~From
$^{66}$Ga(EC).  \\
1056.8  &  0.00164     &   $^{20}$F(1057--g.s.)  & 1056.8 & $a$ ~~~From
$^{20}$O($\beta^-$). \\
1063.4  &    0.0079  &      $^{207m}$Pb(1064--g.s.) & 1063.7 & $s$  \\
1077.9   &  0.0037    &   $^{68}$Zn(1077--g.s.) + L  & 1078.6 & $a$ ~~~From
$^{68}$Ga(EC).  \\
1086.8 &    0.0056  &     $^{68}$Zn(1077--g.s.) + K & 1087.1 & $a$ ~~~From
$^{68}$Ga(EC). \\
1107.9   & 0.0128     &  $^{69}$Ga(1107--g.s.) + L & 1108.1 & $a$ ~~~From
$^{69}$Ge(EC).  \\
                &    &   $^{71}$Ge(1096--g.s.) + K & 1106.6 & $a$ ~~~From 
$^{71}$As(EC).  \\
1117.2    &  0.086  &    $^{69}$Ga(1107--g.s.) + K & 1117.1 & $a$ ~~~From
$^{69}$Ge(EC).   \\
  &                     & $^{65}$Cu(1116--g.s.) + L & 1116.6 & $a$
~~~From $^{65}$Zn(EC).  \\
      &           &        $^{65}$Cu(1116--g.s.)  & 1115.5 & $s$     \\
1124.6  &    0.041 &  $^{65}$Cu(1116--g.s.) + K & 1124.5 & $a$ ~~~45\% 
buildup on $^{65}$Zn $\tau_{1/2}$. \\
     &           &         $^{21}$Ne(2867--1746) & 1121   & $s$    \\
1130.4  &    0.020    &       $^{26}$Mg(2938--1809)  & 1129.7 & $s$   \\
1157.0  &    0.0128  &      $^{44}$Ca(1157--g.s.) & 1157.0 & $a$ ~~~From
$^{44}$Sc($\beta^+$).  \\
    &   &                   $^{214}$Po(2699--1543) & 1155.6 & $r$ ~~~$^{238}$U
series.  \\
1173.2  &    0.0126  &      $^{60}$Ni(1173--g.s.) & 1172.9 & $s$
~~~From $^{60}$Co($\beta^-$). \\
1227.2  &    0.0021     &      $^{42}$Ca(2752--1525)  & 1227.7 & $a$ ~~~From
$^{42}$Sc($\beta^+$). \\  
1237.5  &    0.011 &       $^{214}$Po(1847--609) & 1238.1 & $r$ ~~~$^{238}$U
series.  \\
        &     &                $^{56}$Fe(2085--847) & 1238.3 & $s$ \\
1245.8  & 0.00078  & $^{56}$Fe(2085--847) + K  & 1245.5 & $a$ ~~~From 
$^{56}$Co(EC). \\   
   &   &  $^{14}$C(7341--6094) & 1248 &  $s$   \\
1274.2  &    0.027 &  $^{22}$Ne(1275--g.s.) & 1274.5 & $s$,$a$ ~From 
$^{22}$Na($\beta^+$). \\
1284.1  &  0.0023     &     $^{47}$Ti(1444--159)  & 1284.9  & $s$  \\
1303.7  & 0.00130     &     $^{47}$Ti(2749--1444)  & 1304.6  &  $s$ \\
1311.7   &     0.013  &      $^{48}$Ti(2296--984) & 1312.1 & $a$ ~~~From 
$^{48}$V($\beta^+$).  \\
1316.4  &  0.0064      &     $^{48}$Ti(2296--984) + K & 1317.1 & $a$ ~~~From 
$^{48}$V(EC).  \\
1326.6   &  0.0049  &      $^{63}$Cu(1327--g.s.) & 1327.0 & $s$  \\
1332.3    & 0.0112   &     $^{60}$Ni(1333--g.s.) & 1332.5 & $s$
~~~From $^{60}$Co($\beta^-$). \\ 
1346.5  &     0.0087 &     $^{69}$Ga(1337--g.s.) + K & 1347.0 & $a$ ~~~From
$^{69}$Ge(EC).   \\
      &           &          $^{24}$Na(1345--g.s.)  & 1344.7 & $s$   \\
      &           &          $^{64}$Ni(1346--g.s.)  & 1345.8 & $s$   \\
1368.5  &    0.126  &      $^{24}$Mg(1369--g.s.) & 1368.6 & $a$,$s$ ~From
$^{24}$Na($\beta^-$).  \\
1393.4   &  0.00166     &    $^{21}$Ne(1746--351)  & 1396 & $s$   \\
1399.7  &  0.0042   &   $^{22}$Na(1984--583) & 1401 & $s$      \\
1407.5   &   0.0070   &    $^{214}$Po(2017--609)    &  1408.0 & $r$
~~~$^{238}$U series. \\
1412.8     &  0.0049    &         $^{63}$Cu(1412--g.s.)  & 1412.1 & $s$   \\
1433.1  &   0.0141 &  $^{52}$Cr(1434--g.s.) & 1434.1 & $s$,$a$ ~From
$^{52}$Mn(EC).  \\
   &     &        $^{234}$U(1435--g.s.) & 1435.4 & $r$ ~~~$^{238}$U series. \\
   &     &      ? $^{27}$Al(4410--2982)  & 1428.1  & $s$  \\
1460.4  &   0.040  & $^{40}$Ar(1461--g.s.) & 1460.8 & ~~~~~$^{40}$K
calibration source. \\
1480.4  &    0.0028 &   $^{65}$Cu(1482--g.s.) & 1481.8 & $s$    \\
1508.9   &   0.0024  & $^{24}$Na(1512--g.s.) & 1512.3 &  $s$  \\ 
     &          &            $^{214}$Po(2119--609) & 1509.2 & $r$
~~~$^{238}$U series.  \\
1525.7  &   0.0086  &       $^{22}$Na(1528--g.s.)  & 1528.1 &  $s$  \\
    &  &       ? $^{42}$Ca(1525--g.s.) & 1524.7 & $a$ ~~~From 
$^{42}$Sc($\beta^+$).  \\
1546.9  & 0.00174  &             $^{63}$Cu(1547--g.s.) & 1547.0 & $s$    \\
1553.0  &  0.0025 &   $^{50}$Ti(1554--g.s.) & 1553.8 & $a$ ~~~From
$^{50}$V(EC). \\ 
1571.4      & 0.0016      &  $^{201}$Pb(2507-990) & 1570.8 & $s$   \\
1649.6   & 0.0018     &  $^{26}$Al(2069--417) & 1652 & $s$ ~~~2 lines \\
1697.0   &   0.0022     &     $^{27}$Mg(1698--g.s.) & 1697.9 & $s$   \\
1721.5 & 0.0142 & $^{27}$Al(2735--1014)  & 1720.3 & $s$ ~~~Broad\\
1726.4      & 0.0075      &          $^{214}$Po(1730--g.s.) & 1729.6 & $r$
~~~$^{238}$U series.  \\
     &        &      $^{24}$Na double esc. & 1732.0 & $s$  \\
1758.4   &   0.0062    &     $^{214}$Po(1764--g.s.)  & 1764.5 & $r$  
~~~$^{238}$U series. \\
     &       &             $^{25}$Mg(2738--975)  & 1762.9 & $s$    \\
1778.5  &    0.0061  &     $^{28}$Si(1779--g.s.)  & 1778.9 & $a$ ~~~From 
$^{28}$Al($\beta^-$). \\
1792.4   &  0.00104  &     $^{25}$Mg(3405--1612)  & 1793.4  &  $s$ \\
1809.0  &    0.049  &       $^{26}$Mg(1809--g.s.) & 1808.6 & $s$,$a$ ~From
$^{26}$Na($\beta^-$). \\    
                &    &    $^{56}$Fe(2658-847)  &1810.8 & $a$ ~~~From 
$^{56}$Mn($\beta^-$).  \\
1844.8   &   0.00176 &   $^{214}$Po(1847--g.s.)  & 1847.4 & $r$ ~~~$^{238}$U
series.  \\
1861.0    & 0.00132     &    $^{63}$Cu(1861--g.s.)  & 1861.3 & $s$    \\
1871.2  & 0.00031  & $^{24}$Na(3217--1345)  & 1872.0  &  $s$ \\
1882.9     &  0.0016   &         $^{15}$N(7155--5271)  & 1884.8 & $s$  \\
1902.2  &  0.0017  & $^{24}$Na(3745--1846)  &  1899.0  & $s$ \\
    &     &   $^{69}$Ga(1892--g.s.) + K  & 1901.8 & $a$ ~~~From
$^{69}$Ge(EC).  \\
1947.6   &   0.00115   &     $^{69}$Ga(1924--g.s.) + K  & 1934.4 & $a$ ~~~From
$^{69}$Ge(EC).  \\
1966.0    &  0.00115      &    $^{25}$Mg(1965--g.s.)  & 1964.5 & $s$   \\
1981.6   &     0.0056  &      $^{18}$O(1982--g.s.)  & 1982.0 & $s$  \\
2032.3   &  0.0015    & $^{69}$Ga(2024--g.s.) + K  & 2034.0 & $a$ ~~~From
$^{69}$Ge(EC). \\
2055.1  &  0.0012   &         $^{25}$Mg(5462--3405)  & 2056.4 & $s$   \\
        &            &        $^{23}$Mg(2051--g.s.)  &  2051 &  $s$  \\
2081.3  &  0.0014    &     $^{22}$Ne(3357--1275)   & 2082.5 & $s$,$a$  ~~From
$^{22}$F($\beta^-$).  \\
2129.5   &   0.0022     &    $^{11}$B(2124.7--g.s.)  & 2124.5 & $s$   \\  
    &     &      $^{26}$Mg(3941--1809)  &  2132.0  &  $s$ \\
2166.8   & 0.0029   &     $^{22}$Ne(5523--3357)  & 2165.9 & $s$,$a$ ~From
$^{22}$F($\beta^-$).  \\
      &    &                 $^{38}$Ar(2167--g.s.) & 2167.4 &  $a$  ~~~From
$^{38}$K($\beta^+$). \\
2183.5  &  0.00085  &  $^{6}$Li(2186--g.s.)  & 2186 & $s$  \\
    &         &             $^{17}$O(3055--871)    & 2184.5 &  $a$ ~~~From 
$^{17}$N($\beta^-$).   \\
%2209.5 & 0.0275 &  $^{27}$Al(2211--g.s.)  & 2211.0 & $s$ ~~~Broad \\ 
2209.5 & 0.0275 &  $^{27}$Al(2211--g.s.)  & 2211.0 & $s$
~~~Kinematically broadened. \\
   &  &  $^{214}$Po(2204--g.s.)  & 2204.2  &  $r$ ~~~$^{238}$U series.  \\
2220.3 & 0.0171 & $p$($n$,$\gamma$)$d$ direct capture  & 2223 & $n$\\
2241.3    &  0.0051  &  $^{24}$Mg single esc. & 2243.0 & $a$ ~~~From 
$^{24}$Na($\beta^-$).    \\
2312.0    &    0.0030  &     $^{14}$N(2313--g.s.) & 2312.6 & $s$   \\
2436.4  &  0.00077   &     $^{21}$Ne(2789--351)  & 2438  &  $s$  \\
2508.0  &   0.0109    &  $^{26}$Mg(4318--1809)   & 2509.6 &  $s$ ~~~Broad
(kinematics?).    \\
2614.9  &   0.0139  &   $^{208}$Pb(2614.6--g.s.)  & 2614.5 & $r$
~~~$^{232}$Th series.  \\     
     &             &     $^{20}$Ne(4248--1634) & 2613.8 &  $s$   \\
2753.7  &    0.028   &      $^{24}$Mg(4123--1369) & 2754.0 & $a$,$s$ ~From
$^{24}$Na($\beta^-$).  \\
2761.4  &      0.0045   &  $^{66}$Zn(2752--g.s.) + K & 2761.6 &  $a$ ~~~From
$^{66}$Ga(EC).  \\
2981.8 & 0.0038 & $^{27}$Al(2982--g.s.) & 2981.8 & $s$ ~~~Broad.\\
     &     &             $^{23}$Na(2982--g.s.) & 2981.9 & $s$    \\
%3003.1 & 0.0120 &   $^{27}$Al(3004--g.s.) & 3004.0 &  $s$ ~~~Broad.\\
3003.1 & 0.0120 &   $^{27}$Al(3004--g.s.) & 3004.0 &  $s$ ~~~Broad
(kinematics?).\\ 
3336.0      &  0.00175   &   $^{13}$C single esc. & 3342.2 & $s$ \\
%3404 & 0.009 & $^{12}$C double esc.  & 3416.0 & $s$ ~~~Broad.\\
3404 & 0.009 & $^{12}$C double esc.  & 3416.0 & $s$ ~~~Kinematically
broadened.\\ 
3683.9    &  0.0015   &    $^{13}$C(3684--g.s.) & 3683.9 & $s$  \\
3800.4  &   0.0029  &    $^{66}$Zn(3792--g.s.) + K & 3801.2 & $a$ ~~~From
$^{66}$Ga(EC). \\
3853.5  &     0.00171   &  $^{13}$C(3854--g.s.) & 3853.2 & $s$  \\
      &    & ? $^{24}$Mg(5235--1369)  & 3866.2 & $s$ \\
%3920 & 0.012 & $^{12}$C single esc.  & 3927.0 & $s$ ~~~Broad.\\
3920 & 0.012 & $^{12}$C single esc.  & 3927.0 & $s$ ~~~Kinematically
broadened.\\ 
4090.0   &   0.0017  & $^{66}$Zn(4086)--g.s.) + K & 4095.9 & $a$ ~~~From 
$^{66}$Ga(EC).  \\
4122.3 &  0.00094  & $^{24}$Mg(4122.6--g.s.) & 4122.7 & $s$,$a$ ~Sum
peak. From $^{24}$Na($\beta^-$). \\
4246.8  &  0.0017   &      $^{15}$N double esc. & 4247.2 & $s$  \\
4306.3  &   0.0013  & $^{66}$Zn(4296--g.s.) + K  & 4305.5 & $a$ ~~~Both from
 $^{66}$Ga(EC).    \\
    &        &    $^{66}$Zn(4296--g.s.) + L  & 4297.1 & $a$ \\
%4431 & 0.015 & $^{12}$C(4439--g.s.) & 4438.0 & $s$ ~~~Broad. \\
4431 & 0.015 & $^{12}$C(4439--g.s.) & 4438.0 & $s$ ~~~Kinematically
broadened.\\ 
4588.9  &   0.0015   &  $^{14}$N single esc. & 4593.8 &  $s$  \\
4757.3   &     0.0017 &     $^{15}$N single esc. & 4758.2 & $s$  \\
5106.0   &    0.0037  &       $^{16}$O double esc.  & 5106.6 & $s$   \\
   &    &   $^{14}$N(5106--g.s.)   & 5104.9  &  $s$ \\
5269.3    &    0.0026   &    $^{15}$N(5270--g.s.) & 5269.2 &  $s$
\\
5289. & 0.0021 & $^{15}$N(5298--g.s.) & 5297.8  & $s$ ~~~Broad ??? \\
5298.0   &  0.00048  &  $^{15}$N(5298--g.s.) & 5297.8  &  $a$ ~~~From 
$^{15}$C($\beta^-$)  \\
5420. & 0.0013 & ? & ? & ? ~~~Broad \\
5617.8    &    0.0044   &     $^{16}$O single esc.   & 5617.6 & $s$    \\
6128.7    &    0.0052   &      $^{16}$O(6129--g.s.) & 6128.6 &  $s$ \\
6242.1 & 0.00006 & ? & ? & ? \\
6723.6 & 0.00005 & ? & ? & ? \\
\enddata
\tablenotetext{a}{ Key to reaction types: $a$ --- $\beta$-decay after
activation; $s$ --- spallation followed by prompt de-excitation; 
$n$ --- ($n$,$\gamma$) or ($n$,$n'$) followed by prompt de-excitation;
$r$ --- natural radioactivity.}

\tablenotetext{b}{ All energies in keV.}
\tablenotetext{c}{ K and L are atomic sub-shell binding energies in 
cases of electron capture (EC).}

\end{deluxetable}

%--- class 2 table ---------------------------------------------------

\begin{deluxetable}{lclcl}
\tabletypesize{\scriptsize}
\tablecaption{Blends containing unidentified lines. \label{class2_ids}
}

\tablewidth{0pt}
\tablehead{
\colhead{TGRS line} & \colhead{Count rate} & \colhead{Transition ID} &
\colhead{Nominal} & \colhead{Comment\tablenotemark{a}} \\ 
\colhead{energy, keV} & \colhead{s$^{-1}$} & \colhead{(levels,
keV)\tablenotemark{b}} & \colhead{energy, keV} & \colhead{} 
}

\startdata

63.7     &    0.35       &   ? $^{65m}$Ni(63--g.s.) & 63.4 & $s$  \\
75.6  & 0.24  &  ? $^{61}$Ni(67--g.s.) + K & 75.3  &  $a$ ~~~From
$^{61}$Cu(EC). \\ 
     &       &    ? $^{66}$Ga(66--g.s.) + K & 75.5 &  $a$ ~~~From
$^{66}$Ge(EC). \\
100.2   &    0.45     &    $^{67}$Zn(93--g.s.) + K  & 103.0 & $a$ ~~~From 
$^{67}$Ga(EC).  \\
109.6   &    0.22     &  $^{19}$F(110--g.s.)  & 109.9 & $s$  \\
246.5  &   0.225  &  $^{214}$Bi(295--53)  & 242.0 & $r$ ~~~$^{238}$U
series. \\   
     &         &      $^{69}$Ga(1107--872) + K & 245.2 & $a$ ~~~From 
$^{69}$Ge(EC). \\  
264.9  &  0.0031   &  ? $^{75}$As(265--g.s.)   & 264.7  & $s$ ~~~Weak \\
366.1   &    0.0055  &        ? $^{73}$Ge(364--67--13--g.s.)   & 364.3 & $n$ 
~~~Three-step transition.   \\
   &       &       ? $^{200}$Hg(368--g.s.) & 367.9  & $a$ ~~~From
$^{200}$Tl(EC). \\
384.3   &    0.0088     &      $^{62}$Cu(426--41) & 385.3 & $s$  \\
412.8    &  0.0082    &          $^{228}$Th(1432--1023)  & 409.5 & $r$ 
~~~$^{232}$Th series. \\
    &      &  $^{54}$Fe(2949--2538)   & 411.4 & $a$ ~~~From 
$^{54}$Co($\beta^+$). \\
    &     &  ? $^{198}$Hg(412--g.s.) & 411.8 & $a$ ~~~From $^{198}$Tl(EC). \\ 
557.4   &    0.0056    & ? $^{62}$Cu(548--g.s.) + K & 557.3 & $a$ ~~~From
$^{62}$Zn(EC).   \\
569.5  &   0.0183 &    $^{207}$Pb(570--g.s.) & 569.6 & $s$  \\
617.2    & 0.0031   &       $^{43}$Ca(990--373) & 617.5 & $a$ ~~~From 
$^{43}$K($\beta^-$).   \\
627.9    &  0.0043   &  $^{63}$Zn(627--g.s.) + L & 628.3 & $a$ ~~~From 
$^{63}$Ga(EC). \\
   &   &  ? $^{201m}$Pb(629--g.s.)  & 629.1 &  $a$ ~~~From $^{201}$Bi(EC). \\
635.2   &   0.0056    &   $^{63}$Zn(637--g.s.) + L  & 638.2 &  $a$ ~~~From
$^{63}$Ga(EC).  \\
  &  &   $^{63}$Zn(627--g.s.) + K  & 636.8 &  $a$ ~~~From $^{63}$Ga(EC).  \\
    &              &       ? $^{71}$Ge(831--525--198)  & 633.1 &  
$n$ ~~~Two-step transition. \\
645.8   &    0.0032     &    $^{63}$Zn(637--g.s.) + K  & 646.0 &  $a$ ~~~From
$^{63}$Ga(EC).  \\
703.2   &    0.082    &    ? $^{53m}$Fe(3040--2339) & 701.1 & $a$ ~~~From
$^{53}$Co($\beta^+$).  \\
      &        &              $^{234}$U(1554--852)  & 701.6 & $r$ ~~~$^{238}$U
series. \\
777.9  &      0.0056 &        $^{228}$Th(1168--396) & 772.3 & $r$
~~~$^{232}$Th series.  \\
787.2  &       0.0073    &  ?  $^{201}$Pb(1415--629) & 786.4 &  $a$ ~~~From
$^{201}$Bi(EC). \\
                  & &         $^{234}$U(786--g.s.) & 786.3 & $r$ ~~~$^{238}$U
series.   \\
826.2    &   0.0036     &   ? $^{60}$Ni(2159--1333)  & 826.1 & $a$ ~~~From
$^{60}$Cu($\beta^+$).\\
     &          &       $^{20}$F(823--gs)  &  822.7 & $s$  \\
851.1   &  0.0077  &  ?    &    & $s$ ~~~Energy from simulation. \\
947.8  &  0.0054 & ? $^{40}$Ar single esc. & 949.8 & ~~~~~From
$^{40}$K calibration source.  \\
1005.9 & 0.0033  & $^{26}$Al(3074--2069)   & 1004.1  & $s$  \\
   &     &   ? $^{48}$V(2062--1056)  & 1006.3 &  $s$  \\
1097.7  & 0.0044  & $^{228}$Th(1153--58) & 1095.7 & $r$ ~~~$^{232}$Th
series. \\ 
 & &                       $^{208}$Pb(3708--2615) & 1093.9 &  $r$ 
~~~$^{232}$Th series. \\
1120.6 & 0.020 &  $^{46}$Ti(2010--889) & 1120.5  & $a$ ~~~From 
$^{46}$Sc(EC). \\
   &   &  $^{21}$Ne(2867--1746) & 1121  & $s$ \\
1163.5  &       0.0070 &     $^{44}$Ca(1157--g.s.) + K & 1161.1 & $a$ ~~~From
$^{44}$Sc(EC).  \\
     &              &    $^{62}$Ni(2336--1173) & 1163.4 & $s$    \\
1190.7   &   0.0040  &    $^{61}$Ni(1185--g.s.) + K & 1193.6 & $a$ ~~~From
$^{61}$Cu(EC).  \\
      &           &           $^{234}$U(1237--43)   & 1193.8 & $r$
~~~$^{238}$U series. \\
1206.9  &    0.00131    &       ? $^{200}$Hg(1574--368)  & 1205.7 &
$a$ ~~~From $^{200}$Tl(EC). \\
1222.2  &  0.0029     &  ? $^{24}$Na(2563--1345)  & 1218.1 &  $s$ \\
1337.6  &  0.0013  &  $^{69}$Ga(1337--g.s.) + L  & 1338.0  & $a$ ~~~From 
$^{69}$Ge(EC). \\
1378.0  &    0.0051     &       $^{214}$Po(1378--g.s.) & 1377.7 & $r$ 
~~~$^{238}$U series. \\
      &           &         $^{57}$Co(1378--g.s.) & 1377.6 & $a$ ~~~From
$^{57}$Ni(EC).  \\   
1439.2  &  0.0064  &   $^{52}$Cr(1434--g.s.) + K  & 1440.1 & $a$  ~~From
$^{52}$Mn(EC).    \\
1609.6  &   0.0063  &   $^{25}$Mg(1612--g.s.)  & 1611.7 & $s$,$a$ ~From 
$^{25}$Na($\beta^-$). \\
1632.1   &     0.032  &    $^{20}$Ne(1634--g.s.)  & 1633.6 & $a$,$s$ 
~From $^{20}$F($\beta^-$).  \\
  &    &    $^{23}$Na(2076--440) & 1636.0 & $s$   \\
1641.2  &  0.0015  & ? $^{38}$Ar(3810--2167)  & 1642.7 &  $s$  \\
1677.9  &  0.0053  &  $^{58}$Fe(1675--g.s.) + K & 1681.8 &  $a$ ~~~From
$^{58}$Co(EC).  \\
1703.9         &   0.0012    &   ? 2211--2223~keV line  & ~~$\sim
1700$ & $s$  \\ 
      &       &             complex single escape  &   &  \\
1711.4   &  0.0024   &    ? $^{73}$As(1796--84)   &  1712.1  &  $s$  \\
1893.3   &   0.0014   & ? $^{63}$Cu(2858--962)  & 1895.6  & $s$  \\
   &   &    ? $^{203}$Pb(2713-820)   & 1893.0 &  $a$ ~~~From
$^{203}$Bi(EC).  \\
2092.2   &  0.0014     &  ? & 2092.5 & $a$  ~~Line energy from simulation.  \\
2103.0   & 0.00164  &  ? 2614~keV single esc. & 2103.5 & $r$,$s$  \\ 
2298.4   &   0.0038    &  $^{15}$N(7566--5271) & 2296.8 & $s$   \\     
   &    &                 $^{11}$B(6743--4445) & 2297.8 & $s$    \\
     &     &     ?   & 2300.5 &    $a$  ~~Line energy from simulation. \\   
2373.8   &   0.00184   &     $^{64}$Zn(2374--g.s.) + L & 2375.5 & $a$ ~~~From
$^{64}$Ga($\beta^+$). \\
2792.2  &  0.00076  & ? $^{21}$Ne(2794--g.s.)  & 2794  &  $s$ ~~~Weak. \\ 
3393.3  &  0.00058    &     $^{66}$Zn(3381--g.s.) + K  & 3391.0 & $a$ ~~~From
$^{66}$Ga(EC).  \\
%\tableline

\enddata

\tablenotetext{a}{ Key to reaction types: $a$ --- $\beta$-decay after
activation; $s$ --- spallation followed by prompt de-excitation; 
$n$ --- ($n$,$\gamma$) or ($n$,$n'$) followed by prompt de-excitation;
$r$ --- natural radioactivity.}
\tablenotetext{b}{ K and L are atomic sub-shell binding energies in 
cases of electron capture (EC).}

\end{deluxetable}

%--- inelastic recoil features --------------------------------------

\begin{deluxetable}{cccccc}
\tabletypesize{\small}
\tablecaption{A comparison of the simulated and the observed count
rate in triangular or ``sawtooth'' shaped features that arise from
inelastic neutron scattering in the Ge detector (see
Figs.~\ref{compare_tgrs_sim_spc_40-800} and
\ref{compare_tgrs_sim_spc_800-1100}). For the two limiting cases of a
low and a high estimate of the underlying continuum, refer to text.
\label{inelast_recoil_table}
}
\tablewidth{0pt}
\tablecolumns{6}
\tablehead{
\colhead{Underlying state,} & \colhead{Energy range,} &
\multicolumn{2}{c}{Low Cont.\ Estimate} & \multicolumn{2}{c}{High
Cont.\ Estimate}\\ \cline{3-4} \cline{5-6}
\colhead{keV} & \colhead{keV} & \colhead{Sim., s$^{-1}$} &
\colhead{Obs., s$^{-1}$} & \colhead{Sim., s$^{-1}$} &
\colhead{Obs., s$^{-1}$}
}
\startdata
596 ($^{74}$Ge)  &  592--686  &  0.66 & 0.6 & 0.37 & 0.28 \\
692 ($^{72}$Ge)  &  687--760  &  0.30 & 0.35 & 0.09 & 0.06 \\
834 ($^{72}$Ge)  &  831--866  &  0.105 & 0.053 & 0.074 & 0.013 \\
1039 ($^{70}$Ge) & 1031--1083 &  0.166 & 0.094 & 0.087 & 0.070 \\
1204 ($^{74}$Ge), 1215 ($^{70}$Ge) & 1200--1260 & 0.153 & 0.109 &
0.070 & 0.061\\
\enddata

\end{deluxetable}

\end{document}